\documentclass[aps,prx,twocolumn,superscriptaddress]{revtex4-1}

\usepackage{amssymb}
\usepackage{amsmath}
\usepackage{bm}
\usepackage{graphicx}
\usepackage{amsfonts}
\usepackage{braket}
\usepackage{subfigure}
\usepackage{hyperref,color}
\usepackage[version=3]{mhchem}

\begin{document}
\title{An explicit formula for high-order sideband polarization by extreme tailoring of Feynman path integrals}

\author{Qile Wu}
\affiliation{Physics Department, University of California, Santa Barbara, California 93106, USA}
\affiliation{Institute for Terahertz Science and Technology, University of California, Santa Barbara, California 93106, USA}

\author{Mark S. Sherwin}
\affiliation{Physics Department, University of California, Santa Barbara, California 93106, USA}
\affiliation{Institute for Terahertz Science and Technology, University of California, Santa Barbara, California 93106, USA}

\date{\today}

\begin{abstract}
High-order sideband generation (HSG), as an analogue of the interband processes in high-harmonic generation (HHG) in solids, is a nonperturbative nonlinear optical phenomenon in semiconductors that are simultaneously driven by a relatively weak near-infrared (NIR) laser and a sufficiently strong terahertz (THz) field. We derive an explicit formula for sideband polarization vectors in a prototypical two-band model based on the saddle-point method. Our formula connects the sideband amplitudes with the laser-field parameters, electronic structures, and nonequilibrium dephasing rates in a highly nontrivial manner. Our results indicate the possibility of extracting information on band structures and dephasing rates from high-order sideband generation experiments with simple algebraic calculations. We also expect our approach to be useful on the quantitative understanding of the interband HHG.
\end{abstract}

\maketitle

\section{Introduction}\label{SEC:introduction}

The recent development of strong laser fields has enabled extensive study of nonperturbative optical responses of crystalline solids in highly nonlinear and nonequilibrium regimes. One celebrated example is high-harmonic generation (HHG), which has been observed in conventional metals~\cite{burnett1977harmonic} and semiconductors~\cite{ghimire2011observation,schubert2014sub,hohenleutner2015real,liu2017high} and serves as an important way to obtain ultraviolet light sources~\cite{garg2016multi,vampa2019characterization}. The realization of HHG in solid crystals has led to a method to probe electronic properties including band structures~\cite{vampa2015all,luu2015extreme,li2020determination,chen2021reconstruction}, Berry curvatures~\cite{luu2018measurement}, topological phases~\cite{bauer2018high,silva2019topological,chacon2020circular,schmid2021tunable,baykusheva2021all,bai2021high,baykusheva2021strong,heide2022probing}, and nonequilibrium dephasing rates of electron-hole coherences~\cite{vampa2014theoretical,du2019probing}. Investigation of HHG in correlated electron systems has also been initiated~\cite{uchida2022high,bionta2021tracking,graanas2022ultrafast}. In semiconductors, HHG contains contributions from intraband and interband processes, which are in general coupled with each other~\cite{golde2008high,golde2009microscopic,avetissian2020many}. The interband process can be understood in a three-step model similar to HHG in atoms~\cite{corkum2007attosecond}. In the first step, an electron-hole pair is created by a strong laser field. In the second step, the electron and hole are accelerated in their respective bands by the same laser field. In the third step, recombination of the electron and hole results in radiation with integer multiples of the fundamental frequencies. The intraband contribution comes from the intraband accelerations of the electron and hole through a nonlinear current~\cite{ghimire2011observation}. We will only discuss the interband processes.

As an analogue of the interband HHG, high-order sideband generation (HSG)~\cite{liu2007high,zaks2012experimental} has also received considerable interest since the last decade~\cite{zaks2013high,banks2013terahertz,langer2016lightwave,banks2017dynamical,valovcin2018optical,langer2018lightwave,borsch2020super,costello2021reconstruction,freudenstein2022attosecond}. HSG occurs in semiconductors when an electron-hole pair is created by a relatively weak near-infrared (NIR) laser with a photon energy $\hbar\Omega$ close to the bandgap $E_{\rm g}$ and then accelerated by a strong terahertz (THz) field with a photon energy $\hbar\omega\ll E_{\rm g}$. Upon recollisions and recombinations of the electron-hole pair, sideband photons of energy $\hbar\Omega+n\hbar\omega$ are emitted, where the sideband index $n$ is an integer~\cite{liu2007high,zaks2012experimental}. In contrast to HHG in semiconductors, intraband and interband processes in HSG are disentangled and separately controlled by two different laser fields. Such simplification has led to a reconstruction of low-energy Bloch wavefunctions of holes in bulk GaAs through a simple algebraic equation~\cite{costello2021reconstruction}. Frequency combs of sidebands with orders $n>100$ (66 sidebands) have been produced from HSG~\cite{valovcin2018optical}. HSG has also played a role in probing Berry curvatures~\cite{banks2017dynamical}, band structures~\cite{borsch2020super}, and electron correlations~\cite{freudenstein2022attosecond}.

Theoretical approaches based on or equivalent to the semiconductor Bloch equations (SBEs)~\cite{lindberg1988effective} have been widely used in the numerical analyses of both the intraband and interband HHG~\cite{golde2008high,golde2009microscopic,golde2011microscopic,vampa2014theoretical,vampa2015semiclassical,schubert2014sub,hohenleutner2015real,mcdonald2015interband,yu2016dependence,luu2016high,jiang2017effect,jiang2018role,floss2018ab,li2019phase,jiang2019crystal,li2020determination,yue2020imperfect,avetissian2020many,chacon2020circular,yue2020structure,yue2021expanded,baykusheva2021strong,chen2021reconstruction,wilhelm2021semiconductor,ngo2021microscopic,chen2021reconstruction,kim2022theory,avetissian2022high,ren2022orientation}. The scattering terms in the SBEs are mostly approximated through a dephasing constant for the interband polarization~\cite{golde2008high,golde2009microscopic,golde2011microscopic,vampa2014theoretical,vampa2015semiclassical,schubert2014sub,hohenleutner2015real,mcdonald2015interband,luu2016high,yu2016dependence,jiang2017effect,jiang2018role,floss2018ab,li2019phase,jiang2019crystal,li2020determination,yue2020imperfect,avetissian2020many,chacon2020circular,yue2020structure,yue2021expanded,baykusheva2021strong,chen2021reconstruction,wilhelm2021semiconductor,ngo2021microscopic,chen2021reconstruction,kim2022theory,avetissian2022high,ren2022orientation}. More details on the scattering effects have also been investigated through the coupling between the density matrix elements and four-point correlations~\cite{langer2018lightwave,borsch2020super,freudenstein2022attosecond}. In the simplest case, the SBEs are solved in the single-electron limit, where the Coulomb interaction between the charge carriers are neglected~\cite{golde2008high,golde2009microscopic,vampa2014theoretical,vampa2015semiclassical,schubert2014sub,hohenleutner2015real,mcdonald2015interband,yu2016dependence,luu2016high,jiang2017effect,jiang2018role,floss2018ab,li2019phase,jiang2019crystal,chacon2020circular,yue2020imperfect,yue2020structure,yue2021expanded,baykusheva2021strong,chen2021reconstruction,ngo2021microscopic,wilhelm2021semiconductor,chen2021reconstruction,kim2022theory,avetissian2022high,ren2022orientation}. Another important aspect is the global gauge symmetry, which has long been ignored in the study of HHG and is paid attention to only recently~\cite{chacon2020circular,yue2020structure,avetissian2020many,yue2021expanded,baykusheva2021strong,ngo2021microscopic,wilhelm2021semiconductor,kim2022theory,avetissian2022high,ren2022orientation}. In fact, to explore the effects of Berry curvatures in HSG, dynamical equations equivalent to the SBEs in the limit of negligible carrier densities and Coulomb interaction have already been used in the forms obeying the global gauge symmetry~\cite{yang2013berry,yang2014giant,yang2015geometric,banks2017dynamical}. A gauge-invariant density-matrix formalism has also been applied in a discussion on the detection of the macroscopic Berry curvature~\cite{virk2011optical}. Theoretical frameworks other than the SBEs in the study of interband HHG include the time-dependent density-functional theory~\cite{otobe2012first,otobe2016high,tancogne2017ellipticity,tancogne2017impact,tancogne2018atomic,floss2018ab,floss2019incorporating,klemke2019polarization,li2019phase,yu2020higher,neufeld2022probing,tancogne2022effect,freeman2022high,yamada2023propagation} and the single-particle time-dependent Schr\"odinger equation~\cite{faisal1997floquet,gupta2003generation,higuchi2014strong,wu2015high,osika2017wannier,du2018multichannel,li2019reciprocal,chen2021reconstruction,li2021huygens}. To gain intuitive pictures of the interband HHG, discussions have been focused on the single-electron limit with the carrier occupations ignored such that the interband polarization can be written in a compact form of Feynman path integrals, which can then be analyzed through the well-established saddle-point method~\cite{salieres2001feynman,vampa2014theoretical,vampa2015semiclassical,mcdonald2015interband,jiang2017effect,osika2017wannier,yue2020imperfect,parks2020wannier,li2020determination,yue2021expanded,li2021huygens}. The three-step model in interband HHG has been extended to include the effects from nonzero Berry curvatures~\cite{yue2020imperfect,yue2021expanded} and imperfect recollisions~\cite{osika2017wannier,parks2020wannier,yue2020imperfect,yue2021expanded,li2021huygens}. A four-step model was also proposed~\cite{li2019reciprocal}. While qualitative understandings of interband HHG have been reached in various aspects, quantitative understandings based on the saddle-point method were initiated just recently~\cite{parks2020wannier}.

The theoretical analyses of HSG were mostly based on either a time-dependent Schr\"odinger equation~\cite{liu2007high,yan2008theory,banks2013terahertz,yang2013berry,yang2014giant,yang2015geometric,xie2013effects}, or a dynamical equation of the interband density matrix elements in the single-electron limit~\cite{crosse2014quantum,crosse2014theory,banks2017dynamical}. Both of these equations are equivalent to the SBEs with negligible carrier occupations and phenomenological dephasing rates. While numerical solutions of SBEs have provided insights on effects from Coulomb interactions in HSG from systems involving strongly bound excitons~\cite{langer2016lightwave,langer2018lightwave,borsch2020super,freudenstein2022attosecond}, analyses in the single-electron limit serve as an important middle stage for investigating more complicated systems and have already led to predictions of many nontrivial emergent phenomena such as dynamical birefringence~\cite{banks2017dynamical}. Similar to the interband HHG, the sideband amplitudes in the single-electron limit were represented by Feynman path integrals, which were analyzed with the saddle-point method~\cite{liu2007high,yan2008theory,yang2013berry,yang2014giant,yang2015geometric,xie2013effects,banks2017dynamical}. Remarkably, agreement between the saddle-point approximation and the full evaluation of the Feynman path integrals can be achieved not only qualitatively but also quantitatively~\cite{yan2008theory,yang2013berry,yang2014giant,yang2015geometric,xie2013effects}. However, from the numerical saddle-point solutions, it is still not fully clear how the laser-field parameters, electronic structures, and nonequilibrium dephasing rates are coded in the sideband amplitudes.

In this paper, we derive an explicit formula for sideband polarization vectors in a prototypical two-band model based on the saddle-point method. To tailor the Feynman path integrals into an explicit algebraic function of the laser-field and material parameters, we notice that, in classical electron-hole recollisions under a linearly-polarized THz field, when the kinetic energy gain of an electron-hole pair is much smaller than their ponderomotive energy in the THz field, the time intervals for the shortest recollision paths lie around the nodes of the THz field, where the THz field is almost linear in time. Our derivation is based on the idea that, for sufficiently large ponderomotive energy in the presence of sufficiently strong dephasing, the shortest recollision paths will dominate such that the THz field can be approximated as near-linear in time in the saddle-point analysis. We call this linear-in-time (LIT) approximation. Our formula connects the sideband amplitudes with the laser-field parameters, electronic structures, and nonequilibrium dephasing rates in a highly nontrivial manner. Our results also indicate the possibility of extracting information about band structures and dephasing rates from HSG experiments with simple algebraic calculations. Owing to the similarity between the interband HHG and HSG, we expect our approach will shed new light on the quantitative understanding of HSG in more complicated systems, as well as interband HHG.

\section{Saddle-point analysis}\label{SEC:saddle_approx}

We start with a saddle-point analysis taking account of only the shortest recollision pathways associated with each sideband in the presence of sufficiently strong dephasing. For simplicity, we convey the idea of the linear-in-time approximation in a prototypical two-band model with zero Berry curvatures and a parabolic energy difference between the conduction and valence bands, $E_{\rm cv}({\bf k})=E_{\rm g}+{\hbar^2k^2}/{(2\mu)}$, where $E_{\rm g}$ is the bandgap, $\hbar$ is the reduced Planck constant, and $\mu$ is the reduced mass of the electron-hole pairs. Under the approximation of free electrons and holes~\cite{liu2007high,yang2013berry,yang2014giant,yang2015geometric}, the $n$th-order sideband polarization vector produced by continuous-wave NIR and THz laser fields can be written as~\cite{banks2017dynamical}
\begin{align}
\mathbb{P}_{n} 
= & \frac{i}{\hbar}\frac{1}{T_{\rm THz}}\int_{0}^{T_{\rm THz}} dt e^{i(\Omega+n\omega )t} 
 \int \frac{d^D{\bf P}}{(2\pi)^D}
 \int_{-\infty}^{t} dt'  {\bf d}^{*}
\notag\\
& \exp
\{
-\frac{i}{\hbar}\int_{t'}^t dt'' (E_{\rm cv}[{\bf k}(t'')]-i\Gamma)
\}
{\bf d}\cdot {\bf E}_{\rm NIR}(t'),
\label{EQ:sideband_amplitude}
\end{align}
which describes a three-step process in HSG as follows. In the first step, an electron-hole pair is created at time $t'$ through the coupling between the interband dipole vector ${\bf d}$ and the electric field of the NIR laser 
${\bf E}_{\rm NIR}(t')={\bf F}_{\rm NIR}e^{-i\Omega t'}$ with frequency $\Omega$, where the rotating wave approximation is used.
In the second step, from time $t'$ to $t$, the electron-hole pair is accelerated by the THz field and accumulates a dynamic phase $(-1/\hbar)\int_{t'}^t dt'' E_{\rm cv}[{\bf k}(t'')]$, where $\hbar{\bf k}(t)=\hbar{\bf P}+e{\bf A}(t)$ is the kinetic momentum with $\hbar{\bf P}$ being the canonical momentum, $e$ the elementary charge, and ${\bf A}(t)$ the vector potential of the THz field. We take the THz field as linearly polarized along x-axis in the form ${\bf F}_{\rm THz}(t)=-\dot{\bf A}(t)=\hat{x}F_{\rm max}\cos(\omega t)$ with frequency $\omega$, and ${\bf A}(t)=-\hat{x}(F_{\rm max}/\omega)\sin(\omega t)$. The constant $\Gamma$ quantifies the dephasing in this step phenomenologically. In the third step, the electron and hole recombine at time $t$ and a sideband with frequency $\Omega+n\omega$ is emitted. Here, $T_{\rm THz}=2\pi/\omega$ is the period of the THz field and $D$ is the dimension of the momentum space. The sideband amplitudes are zero for odd sideband index $n$ because of the inversion symmetry in this two-band model. The sideband polarization vector can be written in the form of Feynman path integrals,
\begin{align}
\mathbb{P}_{n} 
= & {\bf d}^{*}{\bf d}\cdot {\bf F}_{\rm NIR}\frac{i\omega}{\pi\hbar}
\int_{0}^{T_{\rm THz}/2} dt
 \int \frac{d^D{\bf P}}{(2\pi)^D}
\notag\\
& \int_0^{+\infty} d\tau
\exp[{\frac{i}{\hbar}S_n({\bf P},t,\tau)}],
\label{EQ:Feynman_path_int}
\end{align}
where we have introduced a time-duration variable $\tau=t-t'$, and an action
\begin{align}
S_n({\bf P},t,\tau)
& =
n\hbar\omega t
-\int_{t-\tau}^t dt''\frac{\hbar^2}{2\mu}[{\bf P}+\frac{e}{\hbar}{\bf A}(t'')]^2\notag\\
& +i(\Gamma-i\Delta)\tau,
\label{EQ:action_original}
\end{align}
with $\Delta=\hbar\Omega-E_{\rm g}$ being the detuning of the NIR laser. The integral with respect to the recombination time $t$ has been folded to be over half a period of the THz field.

To tailor the Feynman path integrals, we apply the saddle-point method~\cite{liu2007high,yan2008theory,yang2015geometric,xie2013effects} by having a Taylor expansion of the action $S_n({\bf P},t,\tau)$ around the saddle points up to the second-order terms and extending the limits of the integrals to infinities to form Gaussian integrals. In the presence of sufficiently strong dephasing, the amplitude of each sideband is dominantly determined by one shortest recollision pathway within half a period of the THz field. Including only the saddle point $({\bf P}_n,t_n,\tau_n)$ for the $n$th-order sideband that corresponds to the shortest recollision pathway, we obtain the approximate expression (see Appendix~\ref{APP:Saddle_point} for the derivation),
\begin{align}
&\mathbb{P}_{n} 
\approx 
2{\bf C}
\exp[{\frac{i}{\hbar}S^{(t,\tau)}_{\rm sc}(t_n,\tau_n)}]
\notag\\
&\frac{
e^{
-(i/2)[
D\arg(\tau_n)
+\arg({\partial^2_{t_n} S^{(t,\tau)}_{\rm sc}})
+\arg({\partial^2_{\tau_n} S^{(\tau)}_{\rm sc}})
]
}
}
{
\sqrt
{
|
(\omega\tau_n)^D
[{\partial^2_{(\omega t_n)} S^{(t,\tau)}_{\rm sc}}/\hbar]
[{\partial^2_{(\omega \tau_n)} S^{(\tau)}_{\rm sc}}/
\hbar]
|
}
},
\label{EQ:saddle_approximation}
\end{align}
which contains a constant vector
\begin{align}
{\bf C}=\frac{-1}{\hbar\omega}  e^{-i{\pi D}/{4}} (\frac{\mu \omega}{2\pi\hbar})^{{D}/{2}} {\bf d}^{*}{\bf d}\cdot {\bf F}_{\rm NIR},
\end{align}
a semiclassical action,
\begin{align}
&S^{(t,\tau)}_{\rm sc}(t_n,\tau_n)\notag\\
=
&
n\hbar\omega t_n+[i\Gamma+\Delta+U_{\rm p}(\gamma^2(\omega\tau_n)-1)]\tau_n\notag\\
&
+U_{\rm p}\tau_n\alpha(\omega\tau_n)\gamma(\omega\tau_n)\cos[\omega(\tau_n-2t_n)],
\label{EQ:saddle_semiclassical}
\end{align}
and two second-order derivatives,
\begin{align}
\frac{1}{\hbar}\frac{\partial^2 S^{(t,\tau)}_{\rm sc}}{\partial{(\omega t_n)}^2}
=
&
2n
\cot[\omega(\tau_n-2t_n)],
\label{EQ:saddlet_derivatives}
\end{align}
\begin{align}
\frac{1}{\hbar}\frac{\partial^2 S^{(\tau)}_{\rm sc}}{\partial{(\omega \tau_n)}^2}
=
&
\frac{n}{2}[\frac{\alpha^2(\omega\tau_n)+\beta^2(\omega\tau_n)}{\omega\tau_n\alpha(\omega\tau_n)\beta(\omega\tau_n)}+1]\cot[\omega(\tau_n-2t_n)]\notag\\
&
+\frac{n}{2}[\frac{\alpha^2(\omega\tau_n)-\beta^2(\omega\tau_n)}{2\alpha(\omega\tau_n)\beta(\omega\tau_n)}]^2\tan[\omega(2t_n-\tau_n)]\notag\\
&
+\frac{U_{\rm p}}{\hbar\omega}\frac{\alpha^2(\omega\tau_n)-\beta^2(\omega\tau_n)}{\omega\tau_n}.
\label{EQ:saddletau_derivatives}
\end{align}
The semiclassical action $S^{(t,\tau)}_{\rm sc}(t_n,\tau_n)$ is given by evaluating the action $S_n({\bf P},t,\tau)$ at the saddle point $({\bf P}_n,t_n,\tau_n)$, while the second line in Eq.~\ref{EQ:saddle_approximation} arises from the Gaussian quantum fluctuations around the saddle point. 
Here, $U_{\rm p}\equiv{e^2F_{\rm max}^2}/{(4\mu\omega^2)}$ is the ponderomotive energy, and we have introduced the functions $\alpha(x)=\cos(x/2)-\gamma(x)$ and $\gamma(x)=\beta(x)/(x/2)$ with $\beta(x)=\sin(x/2)$. Different from the approximate expressions for sideband amplitudes in Ref.~\cite{yan2008theory},~\cite{yang2015geometric}, and~\cite{xie2013effects}, Eq.~\ref{EQ:saddle_approximation} does not contain square roots of complex numbers, which are not single-valued. The values of ${\bf P}_n$, $t_n$, and $\tau_n$ satisfy the saddle-point equations,
\begin{align}
\int_{t_n-\tau_n}^{t_n} dt''
\frac{\hbar {\bf k}_n(t'')}{\mu}
={\bf 0},  
\label{EQ:x_recollision}
\end{align}
\begin{align}
E_{\rm eh}[k_n(t_n)]-E_{\rm eh}[k_n(t_n-\tau_n)]=n\hbar\omega, \label{EQ:energy_recombination}
\end{align}
\begin{align}
E_{\rm eh}[k_n(t_n-\tau_n)]=i\Gamma+\Delta, \label{EQ:energy_creation}
\end{align}
where $E_{\rm eh}(k)=\hbar^2k^2/(2\mu)$ is the kinetic energy from the relative motion of the electron-hole pairs, and $\hbar{\bf k}_n(t'')=\hbar{\bf P}_{n}+e {\bf A}(t'')$ is the time-dependent kinetic momentum associated with the saddle point. The first saddle-point equation corresponds to the condition that an electron and a hole recombine at the site where they are created. The second and third saddle-point equations are related to energy conservation for the cases with zero dephasing ($\Gamma=0$) and nonnegative detunings ($\Delta\ge0$) upon creation and recombination of the electron-hole pairs, respectively. For the cases with zero dephasing ($\Gamma=0$) and negative detunings ($\Delta<0$), Eq.~\ref{EQ:energy_creation} describes creation of electron-hole pairs through quantum tunneling with a pure imaginary energy~\cite{xie2013effects}. Nonzero dephasing ($\Gamma\neq0$) makes the kinetic energy $E_{\rm eh}[k_n(t'')]$ complex in general during the recollision events. As we will see later in this section, nonzero detunings do not introduce extra obstacles in tailoring the Feynman path integrals, since the sideband polarization vector $\mathbb{P}_n$ depends on the detuning $\Delta$ through an analytic function of the complex variable $i\Gamma+\Delta$. Thus we set $\Delta=0$ in the numerical calculations from here on and postpone the discussion of the effects from nonzero detunings until Section~\ref{SEC:nonzero_detuning}. 
\begin{figure}
	\includegraphics[width=0.47\textwidth]{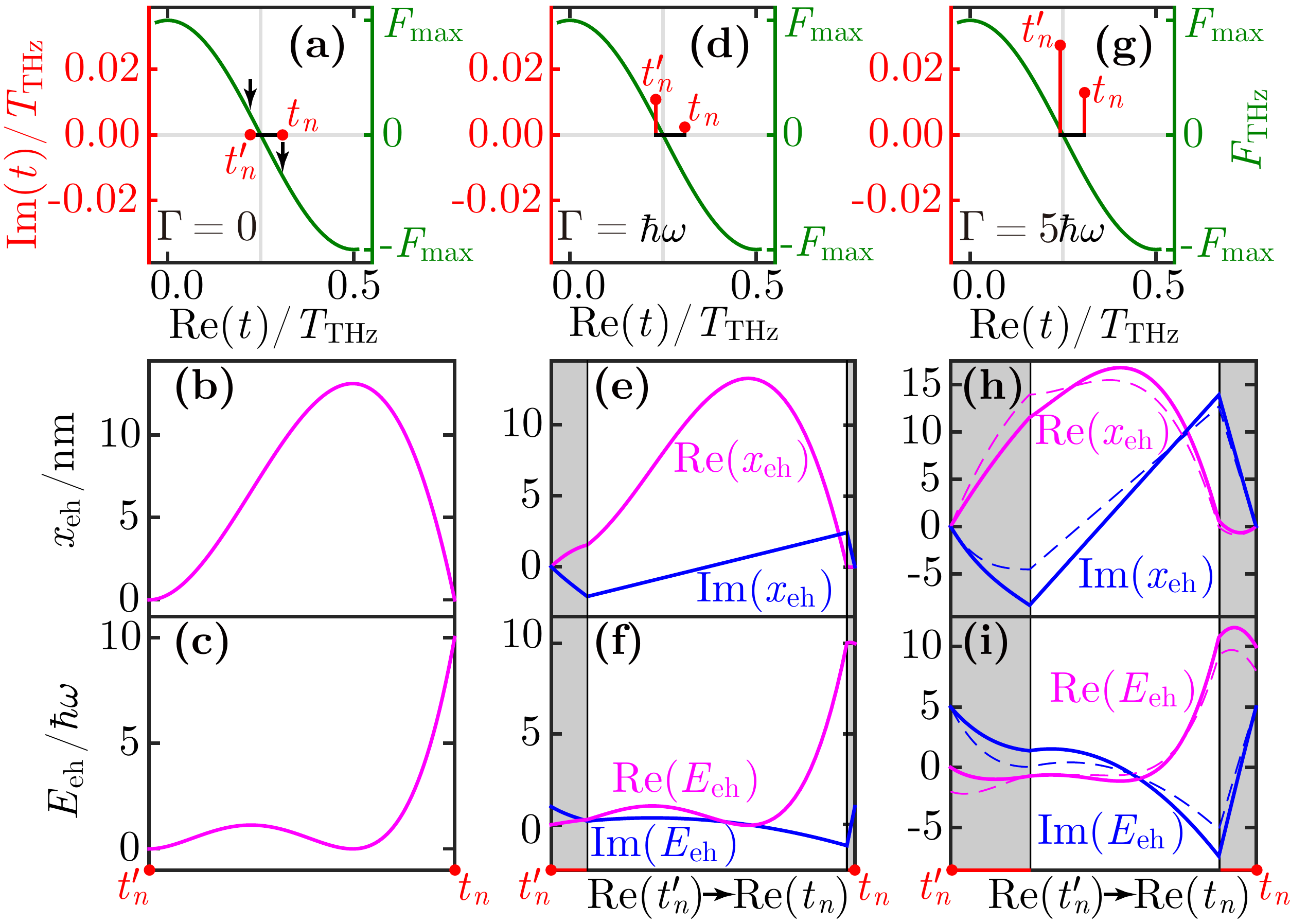}
	\caption{Semiclassical pictures of electron-hole recollisions for the 10th-order sideband. (a) The creation time $t'_n$ and recollision time $t_{n}$ (both real, red dots) in half a period of the THz field $F_{\rm THz}$ (dark green curve) for zero dephasing case ($\Gamma=0$). The THz field is almost linear in time from $t'_n$ to $t_n$ (black arrows). (b) The separation $x_{\rm eh}$ and (c) the kinetic energy $E_{\rm eh}$ (in units of the THz photon energy $\hbar\omega$) of an electron-hole pair along the real time-path from $t'_n$ to $t_n$ (black straight line-segment in (a)). (d) The creation time $t'_n$ and recollision time $t_n$ (both complex, red dots) for the case with dephasing constant $\Gamma=\hbar\omega$. (e) The separation $x_{\rm eh}$ and (f) the kinetic energy $E_{\rm eh}$ of an electron-hole pair along the time path in the complex-time plane, $t'_n\rightarrow{\rm Re}(t'_n)\rightarrow{\rm Re}(t_n)\rightarrow t_n$ (two red straight line-segments parallel to the imaginary-time axis and a black straight line-segment in (d)). Both $x_{\rm eh}$ and $E_{\rm k}$ are complex (magenta and blue curves respectively for the real and imaginary parts). The shaded areas indicate the region where the time is complex. (g), (h), and (i) show results corresponding to (d), (e), and (f), respectively, for the case with a dephasing constant $\Gamma=5\hbar\omega$. In the calculation, we use ponderomotive energy $U_{\rm p}=2\times10^3\hbar\omega$ and $U_{\rm p}/(eF_{\rm THz})$=800 nm. The detuning is set as zero except for the dashed lines in (h) and (i) showing $x_{\rm eh}$ and $E_{\rm eh}$ in the case with dephasing constant $\Gamma=5\hbar\omega$ and detuning $\Delta=-2\hbar\omega$, where the creation time $t'_n$ and recollision time $t_{n}$ are slightly different from those in (g).}
	\label{FIG:semiclassical_picture}
\end{figure}

Using the approximate expression, Eq.~\ref{EQ:saddle_approximation}, one can write the sideband polarization vector $\mathbb{P}_n$ as an explicit function of the laser-field and material parameters on the premise that the explicit forms of $t_n$ and $\tau_n$ are known. However, the saddle-point equations are transcendental in general. To find clues for further approximation, we investigate the semiclassical recollision pictures provided by the saddle-point equations in the special cases where the sideband photon energies are much smaller than the ponderomotive energy ($n\hbar\omega\ll U_{\rm p}$). Fig.~\ref{FIG:semiclassical_picture} shows the time paths of recollisions, electron-hole separation $x_{\rm eh}(t'')=\int^{t''}_{t_n-\tau_n}dt''{\hbar k_n(t'')}/{\mu}$ and kinetic energy $E_{\rm eh}[k_n(t'')]$ for the 10th-order sideband. The ponderomotive energy $U_{\rm p}$ is chosen as $2\times10^3\hbar\omega$, which is a typical value in existing HSG experiments~\cite{costello2021reconstruction}. Fig.~\ref{FIG:semiclassical_picture} (a), (d) and (g) show three time paths corresponding to the shortest recollision pathways within half a period of the THz field (green curves) for the cases with zero detuning and dephasing constants $\Gamma=0,\,\hbar\omega,\, 5\hbar\omega$, respectively. We denote $t'_{n}=t_n-\tau_n$ for the creation time of the electron-hole pairs. Since the kinetic energy $E_{\rm eh}[k_n(t'')]$ and the relative velocity $v_{\rm eh}(t'')={\hbar k_n(t'')}/{\mu}$ are both analytic functions of time, any time path in the complex time plane connecting two fixed time points gives the same dynamic phase and electron-hole separation. For the zero-dephasing case ($\Gamma=0$), the time path can always be chosen as lying on the real-time axis (black line segment in Fig.~\ref{FIG:semiclassical_picture}(a)). This choice corresponds to a classical recollision picture with a real electron-hole separation $x_{\rm eh}$ (Fig.~\ref{FIG:semiclassical_picture}(b)) and a real kinetic energy $E_{\rm eh}$  (Fig.~\ref{FIG:semiclassical_picture}(c)). Remarkably, along such a time path, the THz field is almost linear in time. This approximate linearity remains in the presence of relatively weak dephasing. As shown in Fig.~\ref{FIG:semiclassical_picture} (d) and (g), although the creation time $t'_{n}$ and recollision time $t_{n}$ become complex, the time path can still be chosen as lying around the origin of the complex time plane. We also see that the creation time $t'_{n}$ and recollision time $t_{n}$ are further away from the real-time axis for stronger dephasing. For the weaker-dephasing case ($\Gamma=\hbar\omega$), an imaginary part of the electron-hole separation arises, while the real part resembles the zero-dephasing case (Fig.~\ref{FIG:semiclassical_picture} (e)). As the dephasing gets stronger, the electron-hole separation contains a more significant imaginary part and a real part more distorted from the classical counterpart (Fig.~\ref{FIG:semiclassical_picture} (h)). A similar trend in the kinetic energy is shown in Fig.~\ref{FIG:semiclassical_picture} (f) and (i). As energy conservation is imposed by the saddle-point equations, Eq.~\ref{EQ:energy_recombination} and \ref{EQ:energy_creation}, in each of the cases, the real part of the kinetic energy goes from zero to the sideband offset energy $10\hbar\omega$, while the imaginary part starts and ends at the value of the dephasing constant $\Gamma$.

From the above analysis of the semiclassical recollision pictures, we see that the linear-in-time approximation might be appropriate in solving the saddle-point equations for relatively small sideband index and not too strong dephasing. A more precise statement can be inferred from the saddle-point equations with the canonical momentum $\hbar{\bf P}_n$ eliminated (see Appendix~\ref{APP:Saddle_point}),
\begin{align}
\sin[\omega(\tau_n-2t_n)]
&=
\frac
{n\hbar\omega}
{4U_{\rm p}\alpha(\omega\tau_n)\beta(\omega\tau_n)},
\label{EQ:tau_t_sin}
\end{align}
\begin{align}
\cos[\omega(\tau_n-2t_n)]
&=
\frac
{\alpha^2(\omega\tau_n)+\beta^2(\omega\tau_n)-\xi}
{\alpha^2(\omega\tau_n)-\beta^2(\omega\tau_n)},
\label{EQ:tau_t_cos}
\end{align}
where $\xi=[i\Gamma+\Delta+(n/2)\hbar\omega]/U_{\rm p}$. If the creation time $t'_n=t_n-\tau_n$ and recollision time $t_n$ are located around the node of the THz field such that $|\omega (2t_n-\tau_n)-\pi|,|\omega\tau_n|\ll 1$, there must be ${n\hbar\omega}/{U_{\rm p}}\approx|(\omega\tau_n)^4[\omega (2t_n-\tau_n)-\pi]/12|\ll 1$, and $|\xi|\approx
|(\omega\tau_n)^2[\omega (2t_n-\tau_n)-\pi]^2/8|\ll 1$. In other words, a sufficient condition for the linear-in-time approximation to be valid is that the dephasing constant $\Gamma$, the detuning $\Delta$, and the sideband offset energy $n\hbar\omega$ are all small with respect to the ponderomotive energy $U_{\rm p}$. We will focus on the accuracy of the linear-in-time approximation under this condition.

\begin{figure}
	\includegraphics[width=0.47\textwidth]{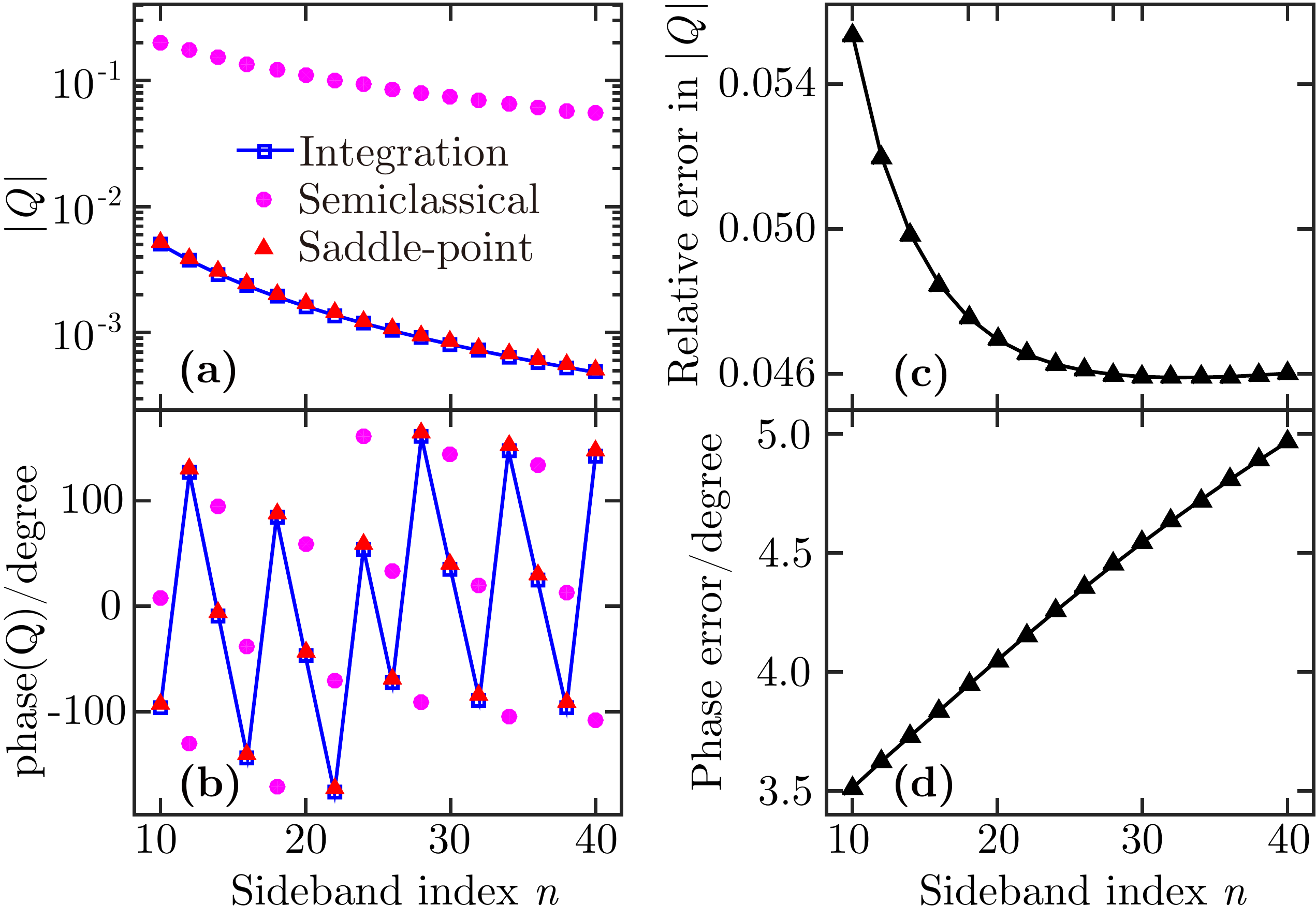}
	\caption{The saddle-point approximation for the dimensionless sideband amplitudes $Q_n$ at relatively low orders of sidebands. (a) and (b) compare respectively the absolute values and phases of $Q_n$ calculated by numerical integration (blue curves) with the results from the saddle-point approximation (red triangles). The magenta dots represent the results solely from the semiclassical propagator $\exp[(i/\hbar)S^{(t,\tau)}_{\rm sc}(t_n,\tau_n)]$. The black curves in (c) and (d) show respectively the relative errors in $|Q_n|$ and the absolute errors in the phases of $Q_n$ in the saddle-point approximation. In the calculation, we use detuning $\Delta=0$, dephasing constant $\Gamma=5\hbar\omega$, and ponderomotive energy $U_{\rm p}=2\times10^3\hbar\omega$.}
	\label{FIG:saddle_point_approx}
\end{figure}

Before exploring the linear-in-time approximation, it is important to know first the accuracy of the saddle-point approximation. To this end, we compare the dimensionless sideband amplitudes $Q_{n}\equiv\mathbb{P}_{n}\cdot{\bf C}/|{\bf C}|^2$ calculated through the saddle-point approximation with the results from numerical integration of the exact expression (see Appendix~\ref{APP:Analytic_calculations}),
\begin{align}
{Q}_{n} 
= & 
i^{n/2-1}
\int_0^{+\infty} \frac{d(\omega\tau) }{(\omega\tau)^{D/2}}
J_{n/2}[\frac{U_{\rm p}}{\hbar\omega}\omega\tau\gamma(\omega\tau)\alpha(\omega\tau)]\notag\\
&
\exp\{i[\mathbb{S}^{(\tau)}(\omega\tau)+n/2]\omega\tau\},
\label{EQ:exact_form}
\end{align}
where $\mathbb{S}^{(\tau)}(\omega\tau)= (i\Gamma+\Delta)/(\hbar\omega)+[U_{\rm p}/(\hbar\omega)][\gamma^2(\omega\tau)-1]$ and $J_n$ is the $n$th-order Bessel function of the first kind. 
We will present numerical results in the main text only for the one-dimensional case ($D=1$). The results are similar for the two- and three-dimensional cases ($D=2,3$) with a linearly-polarized THz field (see Fig.~\ref{FIG:detuning_abs_n40_dimk2}-\ref{FIG:detuning_phase_n40_dimk3} in Appendix~\ref{APP:supplement_figures} for example results regarding the accuracy of the linear-in-time approximation). Fig.~\ref{FIG:saddle_point_approx} shows a comparison for sideband indices from 10 to 40. The ponderomotive energy is chosen as $U_{\rm p}=2\times10^3\hbar\omega$, the same typical value in HSG experiments~\cite{costello2021reconstruction} as in Fig.~\ref{FIG:semiclassical_picture}, and the dephasing constant is set as $\Gamma=5\hbar\omega$. As shown in Fig.~\ref{FIG:saddle_point_approx} (a) and (b), the saddle-point approximation agrees well with the numerical integration for both the absolute values and phases of the sideband amplitudes. We also see that the variations of the dimensionless sideband amplitudes $Q_n$ with respect to the sideband index $n$ closely follow those of the semiclassical propagator, $\exp[(i/\hbar)S^{(t,\tau)}_{\rm sc}(t_n,\tau_n)]$. However, the absolute values of the semiclassical propagator are off by about two orders of magnitude from the numerical integration results (Fig.~\ref{FIG:saddle_point_approx} (a)), while the phases are off by around 100 degrees (Fig.~\ref{FIG:saddle_point_approx} (b)). Therefore, the Gaussian quantum fluctuations are important in determining the sideband amplitudes. To quantify the accuracy of the saddle-point approximation, we compute the relative errors in the absolute values of $Q_n$ and absolute errors in the phases of $Q_n$ with respect to the numerical integration results. As shown in Fig.~\ref{FIG:saddle_point_approx} (c) and (d), within the considered sideband window, the relative errors in the absolute values of $Q_n$ stay around 5\%, and the absolute phase errors go from about 3.5 to 5 degrees.

\begin{figure}
	\includegraphics[width=0.47\textwidth]{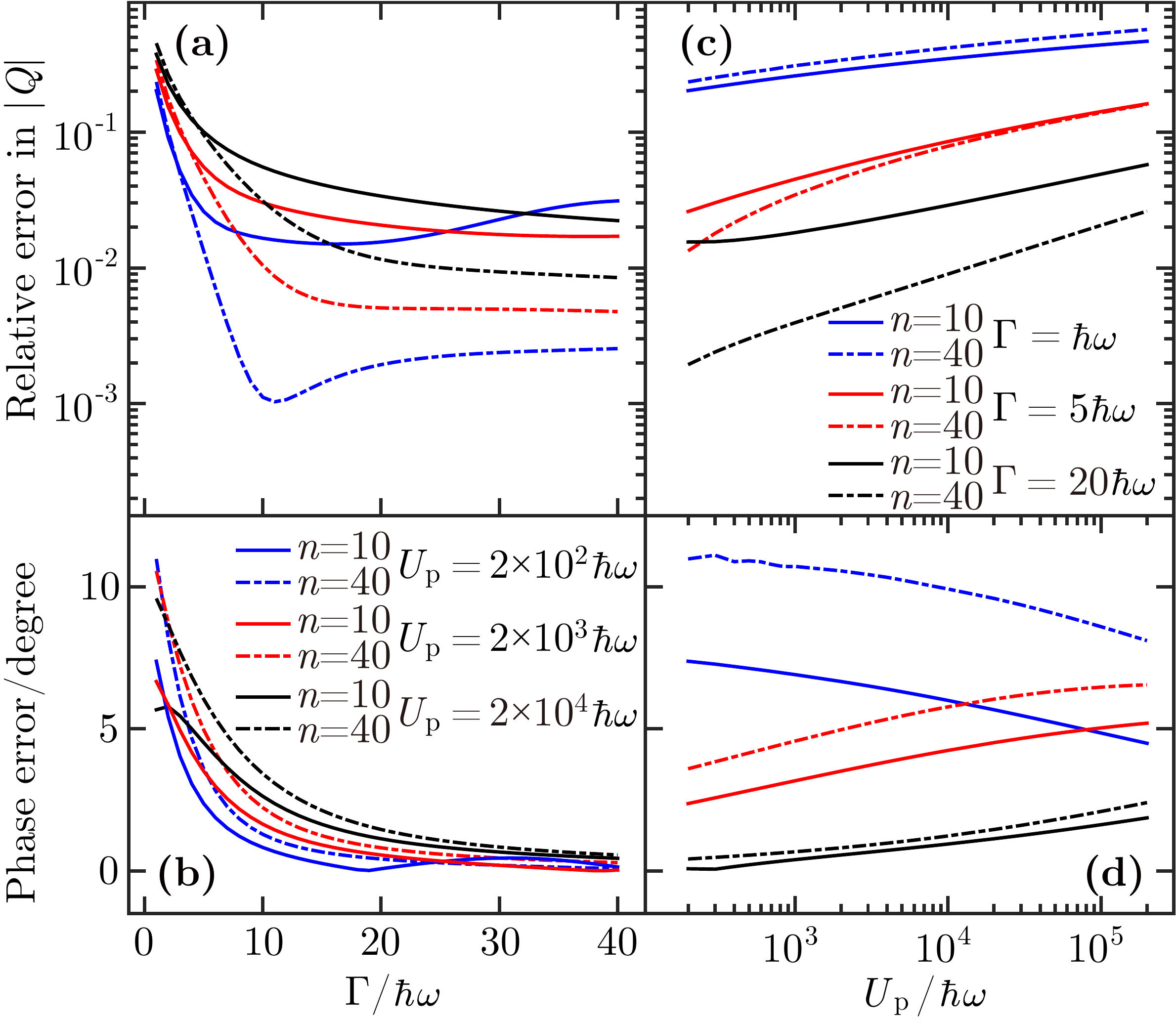}
	\caption{The accuracy of the saddle-point approximation for the dimensionless sideband amplitudes $Q_n$. (a) and (b) show respectively the relative errors in $|Q_n|$ and the absolute errors in the phases of $Q_n$ as functions of the dephasing constant $\Gamma$ with the ponderomotive energy $U_{\rm p}$ fixed at $2\times10^2\hbar\omega$ (blue curves), $2\times10^3\hbar\omega$ (red curves), and $2\times10^4\hbar\omega$ (black curves). (c) and (d) show respectively the relative errors in $|Q_n|$ and absolute errors in the phases of $Q_n$ as functions of the ponderomotive energy $U_{\rm p}$ with the dephasing constant $\Gamma$ fixed at $\hbar\omega$ (blue curves), $5\hbar\omega$ (red curves), and $20\hbar\omega$ (black curves). The results for sideband indices $n=10$ and $n=40$ are plotted as solid and dash-dotted curves, respectively. Zero detunings are used for all cases.}
	\label{FIG:saddle_point_accuracy}
\end{figure}
To have a more systematic view of how the accuracy of the saddle-point approximation varies with the laser-field and material parameters, we first notice that, apart from the sideband index $n$, each dimensionless sideband amplitude $Q_n$ is solely determined by two quantities, a combination of the dephasing constant and detuning, $(i\Gamma+\Delta)/(\hbar\omega)$, and the ponderomotive energy $U_{\rm p}/(\hbar\omega)$, both in units of the THz photon energy $\hbar\omega$. This statement is clear from the exact expression, Eq.~\ref{EQ:exact_form}, and is also valid under the saddle-point approximation (see Eq.~\ref{EQ:saddle_approximation},~\ref{EQ:saddle_semiclassical},~\ref{EQ:saddlet_derivatives},~\ref{EQ:saddletau_derivatives},\ref{EQ:tau_t_sin}, and \ref{EQ:tau_t_cos}). Thus we compute the errors in the dimensionless sideband amplitudes $Q_n$ for sideband indices $n=10$ and $n=40$ over a wide range of dephasing constants and ponderomotive energies around the experimentally accessible values in units of the THz photon energy. Fig.~\ref{FIG:saddle_point_accuracy} (a) and (b) show respectively the relative errors in the absolute values of $Q_n$ and the absolute errors in the phases of $Q_n$ as functions of the dephasing constant $\Gamma$ with the ponderomotive energy $U_{\rm p}$ fixed at $2\times10^2\hbar\omega$ (blue curves), $2\times10^3\hbar\omega$ (red curves), and $2\times10^4\hbar\omega$ (black curves). As a general trend, the relative errors in $|Q_n|$ and the phase errors decrease as the dephasing gets stronger, except for some nonmonotonic behaviors in the cases with relatively small ponderomotive energy (e.g., blue curves in Fig.~\ref{FIG:saddle_point_accuracy} (a) and (b)). Fig.~\ref{FIG:saddle_point_accuracy} (c) and (d) show respectively the relative errors in $|Q_n|$ and the absolute errors in the phases of $Q_n$ as functions of the ponderomotive energy $U_{\rm p}$ with the dephasing constant $\Gamma$ fixed at $\hbar\omega$ (blue curves), $5\hbar\omega$ (red curves), and $20\hbar\omega$ (black curves). For larger ponderomotive energy, the errors are mostly larger in the three selected dephasing cases with the exception of the phase errors in the cases with $\Gamma=\hbar\omega$ (blue curves in Fig.~\ref{FIG:saddle_point_accuracy} (d)). Nonmonotonic variations of the errors with increasing ponderomotive energy are also seen for the relatively low-order sideband in the strong-dephasing cases (e.g., solid curves in Fig.~\ref{FIG:saddle_point_accuracy} (a) and (b)). As for the dependences on the sideband indices, the relative errors in $|Q_n|$ are smaller for higher-order sidebands except for the weak-dephasing cases (blue curves in Fig.~\ref{FIG:saddle_point_accuracy} (c)), while the phase errors are smaller for smaller sideband indices in the three selected cases with weak to moderate dephasing (Fig.~\ref{FIG:saddle_point_accuracy} (d)). Over the whole parameter space investigated, the relative errors in $|Q_n|$ are mostly below 10\% and the phase errors are mostly less than 10 degrees.

The results of the accuracy analysis shown in Fig.~\ref{FIG:saddle_point_approx} and~\ref{FIG:saddle_point_accuracy} can be appreciated by considering the wave nature of the electron-hole pairs in HSG. The electrons and holes are generally not point particles but quantum mechanical objects with wavefunctions of finite widths. As has been discussed in Ref.~\cite{crosse2014quantum}, the centers of an electron and a hole wave packets do not even need to coincide with each other to recombine and generate sidebands. Intuitively, one expects that the recollision processes in HSG can be described by the semiclassical trajectories given by the saddle-point method if the maximum separations of the electron-hole pairs are much larger than the widths of their wavefunctions in real space. The maximum separations are larger for higher sideband indices in the limit of classical recollisions, while a direct calculation of the momentum distributions of the electron-hole wavefunctions indicates that the electron-hole wavefunctions tend to be broader in real space for weaker dephasing and lower-order sidebands. This is consistent with the enhanced accuracy of the saddle-point approximation in Fig.~\ref{FIG:saddle_point_approx} by including the Gaussian fluctuations, and the trends shown in Fig.~\ref{FIG:saddle_point_accuracy} (a) and (b) that the saddle-point approximation tends to be more accurate for relatively higher-order sidebands and relatively strong dephasing. The lower accuracy for larger ponderomotive energy shown in most curves in Fig.~\ref{FIG:saddle_point_accuracy} (c) and (d) could also be attributed to the broader electron-hole wavefunctions in real space. See Appendix~\ref{APP:excursion_vs_wavepacket} for more details.

\section{linear-in-time approximation}\label{SEC:linear_field}

Based on the saddle-point analysis, we now continue tailoring the Feynman path integrals using the linear-in-time approximation. The first task is to obtain explicit forms of the creation time $t'_n=t_n-\tau_n$ and the recollision time $t_n$ from the saddle-point equations. Under the linear-in-time approximation, the THz field strength is approximated by the first-order Taylor polynomial at the node $\omega t=\pi/2$, $F_{\rm THz}(t)=-F_{\rm max}(\omega t-\pi/2)$. To make the mathematics simpler, we define time variables with a tilde to indicate a translation of half a period of the THz field, e.g., $\omega \tilde{t}=\omega t-\pi/2$. The kinetic momentum $\hbar k_n(t)$ satisfies the Newtonian equation of motion
\begin{align}
\hbar \dot{k}_n(t)=-eF_{\rm THz}(t)=eF_{\rm max} \omega \tilde{t},
\end{align}
whose solution can be written as 
\begin{align}
k_n(t)=k_n(t'_n)+\frac{eF_{\rm max}}{2\hbar\omega}[(\omega \tilde{t})^2-(\omega  \tilde{t}'_n)^2].
\label{EQ:kn_t}
\end{align}
Putting this solution into the first saddle-point equation, Eq.~\ref{EQ:x_recollision}, yields
\begin{align}
\omega^2(\tilde{t}_n+2\tilde{t}'_n)(\tilde{t}_n-\tilde{t}'_n)=-\frac{6\hbar\omega}{eF_{\rm max}}k_n(t'_n),
\label{EQ:t_n_k_n_1}
\end{align}
which provides a relation connecting the time variables $t'_n$ and $t_n$ with the kinetic momenta $\hbar k_n(t)$ at $t'_n$ and $t_n$. The solution of $k_n(t)$ at $t_n$ provides another such relation,
\begin{align}
\omega^2(\tilde{t}_n+\tilde{t}'_n)(\tilde{t}_n-\tilde{t}'_n)=\frac{2\hbar\omega}{eF_{\rm max}}[k_n(t_n)-k_n(t'_n)].
\label{EQ:t_n_k_n_2}
\end{align}
The saddle-point equations concerning the energy conservation, Eq.~\ref{EQ:energy_recombination} and~\ref{EQ:energy_creation}, are not affected by the linear-in-time approximation, giving the kinetic momenta $\hbar k_n(t)$ at the creation time $t'_n$ and recollision time $t_n$ through the following equations,
\begin{align}
\frac{\hbar\omega}{eF_{\rm max}}k_n(t'_n)=\pm\sqrt{\frac{i\Gamma+\Delta}{2U_{\rm p}}}\equiv\pm\zeta_0\sqrt{\frac{\hbar\omega}{2U_{\rm p}}},
\label{EQ:k_n_0}
\end{align}
\begin{align}
\frac{\hbar\omega}{eF_{\rm max}}k_n(t_n)=\sqrt{\frac{i\Gamma+\Delta+n\hbar\omega}{2U_{\rm p}}}\equiv\zeta_n\sqrt{\frac{\hbar\omega}{2U_{\rm p}}},
\label{EQ:k_n_c}
\end{align}
where $\zeta_n\equiv\sqrt{(i\Gamma+\Delta)/(\hbar\omega)+n}$. We have fixed the sign of $\hbar k_n(t_n)$ to make it continuously connect with the kinetic momentum in the limit of classical recollisions ($\Gamma=\Delta=0$) at the recollision time. In this paper, a square root of a complex number is defined to have a nonnegative real part. From Eq.~\ref{EQ:t_n_k_n_1},~\ref{EQ:t_n_k_n_2},~\ref{EQ:k_n_0} and~\ref{EQ:k_n_c}, the creation time $t'_n$ and the recollision time $t_n$ can be easily solved as
\begin{align}
\omega \tilde{t}'_n
=
(\frac{2\hbar\omega}{9U_{\rm p}})^{1/4}
\frac{
2\zeta_0-\zeta_n
}
{
\sqrt
{
\zeta_n-\zeta_0
}
},
\label{Eq:t_0_linear}
\end{align}
\begin{align}
\omega \tilde{t}_n
=
(\frac{2\hbar\omega}{9U_{\rm p}})^{1/4}
\frac{
2\zeta_n-\zeta_0
}
{
\sqrt
{
\zeta_n-\zeta_0
}
},
\label{Eq:t_c_linear}
\end{align}
which correspond to a time duration with a positive real part,
\begin{align}
\omega \tau_n
=
(\frac{18\hbar\omega}{U_{\rm p}})^{1/4}
{
\sqrt
{
\zeta_n-\zeta_0
}
}.
\end{align}
To make the imaginary part of $\tau_n$ nonpositive regarding the convergence of the Gaussian integrals in the saddle-point approximation (see Appendix~\ref{APP:Saddle_point}), we have chosen the kinetic momentum $\hbar k_n(t'_n)$ to have a nonpositive real part. These solutions are consistent with the sufficient condition discussed in the last section for the validity of the linear-in-time approximation that the dephasing constant $\Gamma$, the detuning $\Delta$, and the sideband offset energy $n\hbar\omega$ should all be small relative to the ponderomotive energy $U_{\rm p}$. In the limit of classical recollisions ($\Gamma=\Delta=0$), the creation time $t'_n$ and the recollision time $t_n$ satisfy $\tilde{t}_n=-2\tilde{t}'_n$, consistent with the numerical results in Fig.~\ref{FIG:semiclassical_picture} (a).

One can arrive at explicit forms of the sideband amplitudes as functions of the laser-field and material parameters by putting the explicit solutions of $t_n$ and $\tau_n$ into the approximate expression from the saddle-point approximation, Eq.~\ref{EQ:saddle_approximation}. However, the dependences of the sideband amplitudes on the laser-field and material parameters are still far from transparent in such forms. To go further, we expand respectively the semiclassical action $S^{(t_n,\tau_n)}(t_n,\tau_n)$ and the two second-order derivatives in Eq.~\ref{EQ:saddle_approximation} into Taylor series up to the terms of the lowest order in $1/U_{\rm p}$,
\begin{align}
\frac{1}{\hbar}S^{(t,\tau)}_{\rm sc}(t_n,\tau_n)
&=
n\omega t_n
+
i\frac{\Gamma}{\hbar\omega}\omega\tau_n\notag\\
-
\frac{U_{\rm p}}{24\hbar\omega}
&
(\omega\tau_n)^3[
\frac{(\omega\tau_n)^2}{15}
+(\omega \tilde{t}'_n+\omega\tilde{t}_n)^2
],
\end{align}
\begin{align}
\frac{1}{\hbar}\frac{\partial^2 S^{(t,\tau)}_{\rm sc}}{\partial{(\omega t_n)}^2}
=
-\frac{1}{3}(\omega \tau_n)^3
\frac{U_{\rm p}}{\hbar\omega},
\end{align}
\begin{align}
\frac{1}{\hbar}\frac{\partial^2 S^{(\tau)}_{\rm sc}}{\partial{(\omega \tau_n)}^2}
=
\frac{U_{\rm p}}{2\hbar\omega}
(\omega\tau_n)
[
(\omega \tilde{t}'_n+\omega\tilde{t}_n)^2
-
\frac{1}{9}(\omega\tau_n)^2],
\end{align}
which lead to a compact algebraic form for the sideband polarization vectors,
\begin{align}
\mathbb{P}_{n} 
\approx 
&
2i^n{\bf C}
\exp
\{
i[
q_{1/4}(n,\frac{i\Gamma+\Delta}{\hbar\omega})(\frac{\hbar\omega}{U_{\rm p}})^{1/4}
]
\}
\notag\\
&
(\frac{U_{\rm p}}{\hbar\omega})^{\frac{D-2}{8}}
\frac{\exp[-i\arg[q_0(n,\frac{i\Gamma+\Delta}{\hbar\omega})]/2]}
{\sqrt{|q_0(n,\frac{i\Gamma+\Delta}{\hbar\omega})|}},
\label{EQ:algebraic_form}
\end{align}
where
\begin{align}
q_{1/4}(n,\frac{i\Gamma+\Delta}{\hbar\omega})
=
&
(\frac{2}{9})^{1/4}
\frac{4\sqrt{\zeta_n - \zeta_0}}{5}
\notag\\
&
(2\zeta_0^2+\zeta_0\zeta_n+2\zeta_n^2),
\label{EQ:definition_q1o4}
\end{align}
\begin{align}
q_{0}(n,\frac{i\Gamma+\Delta}{\hbar\omega})
=
-\sqrt{32(3\sqrt{2})^D}\zeta_0\zeta_n(\zeta_n-\zeta_0)^{\frac{D+2}{2}}.
\label{EQ:definition_q0}
\end{align}
The factor $i^n$ is related to the initial phase of the THz field. As can be easily seen from Eq.~\ref{EQ:sideband_amplitude}, a phase shift of $\varphi$ in the THz field will result in a phase shift of $n\varphi$ in the $n$th-order sideband. 
\begin{figure}
	\includegraphics[width=0.47\textwidth]{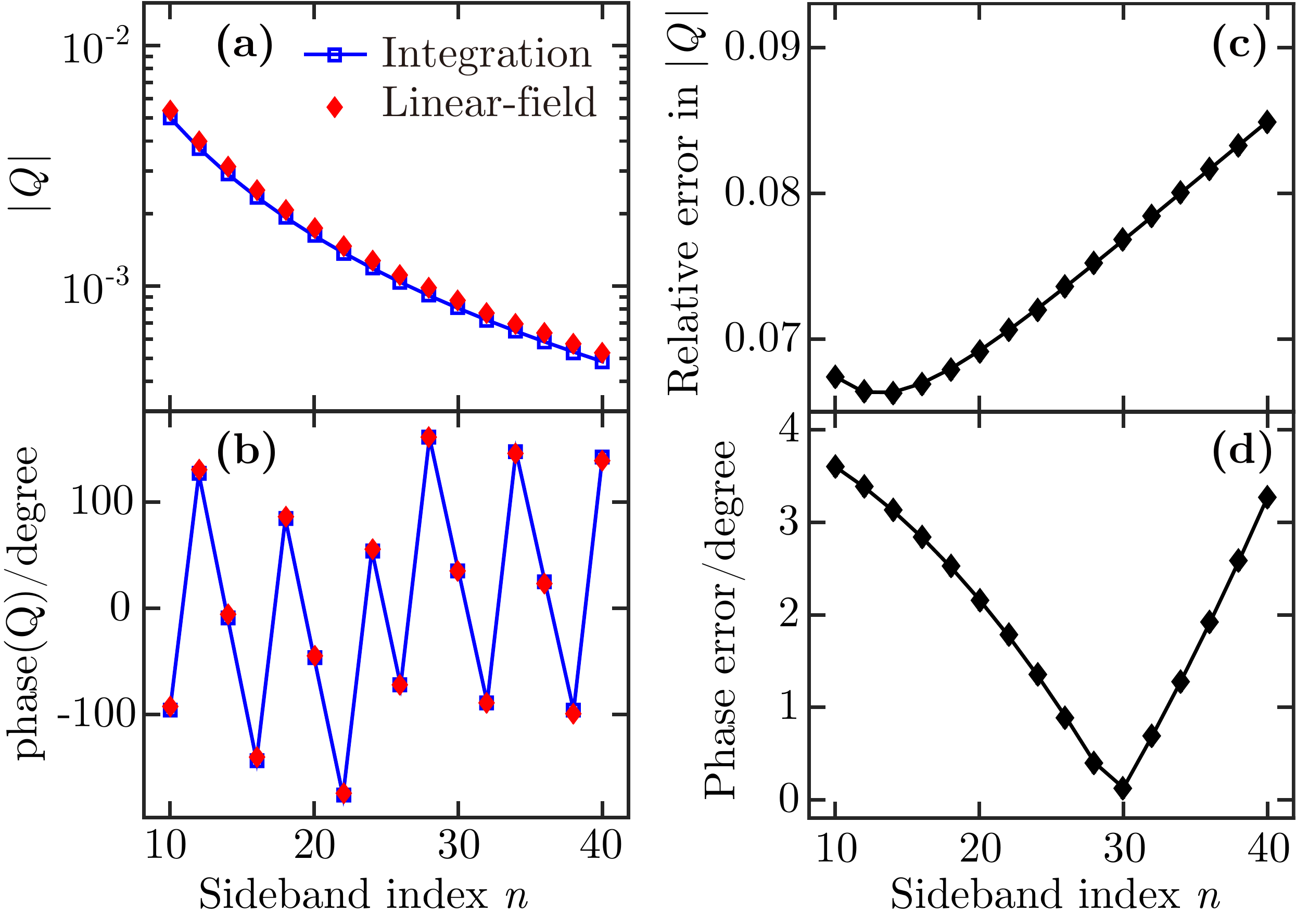}
	\caption{The linear-in-time approximation for the dimensionless sideband amplitude $Q_n$. (a) and (b) compare respectively the absolute values and phases of $Q_n$ calculated by numerical integration (blue curves) to the results from the linear-in-time approximation. The black curves in (c) and (d) show respectively relative errors in $|Q_n|$ and absolute errors in the phases of $Q_n$ in the linear-field approximation. In the calculation, we use detuning $\Delta=0$, dephasing constant $\Gamma=5\hbar\omega$ and ponderomotive energy $U_{\rm p}=2\times10^3\hbar\omega$.}
	\label{FIG:linear_field_approx}
\end{figure}
Fig.~\ref{FIG:linear_field_approx} shows a comparison of the dimensionless sideband amplitues $Q_n=\mathbb{P}_{n}\cdot{\bf C}/|{\bf C}|^2$ calculated from the algebraic form, Eq.~\ref{EQ:algebraic_form}, with the results from numerical integration of Eq.~\ref{EQ:exact_form}. We use the same parameters as in Fig.~\ref{FIG:saddle_point_approx}. As shown in Fig.~\ref{FIG:linear_field_approx} (a) and (b), the algebraic form agrees well with the numerical integration for both the absolute values and phases of the sideband amplitudes. The relative errors in the absolute values of $Q_n$ stay below 9\% (Fig.~\ref{FIG:linear_field_approx} (c)), and the absolute errors in the phases are less than 4 degrees (Fig.~\ref{FIG:linear_field_approx} (d)). The dip in the phase errors at $n=30$ arises from a sign change in the phase differences. 

\begin{figure}
	\includegraphics[width=0.47\textwidth]{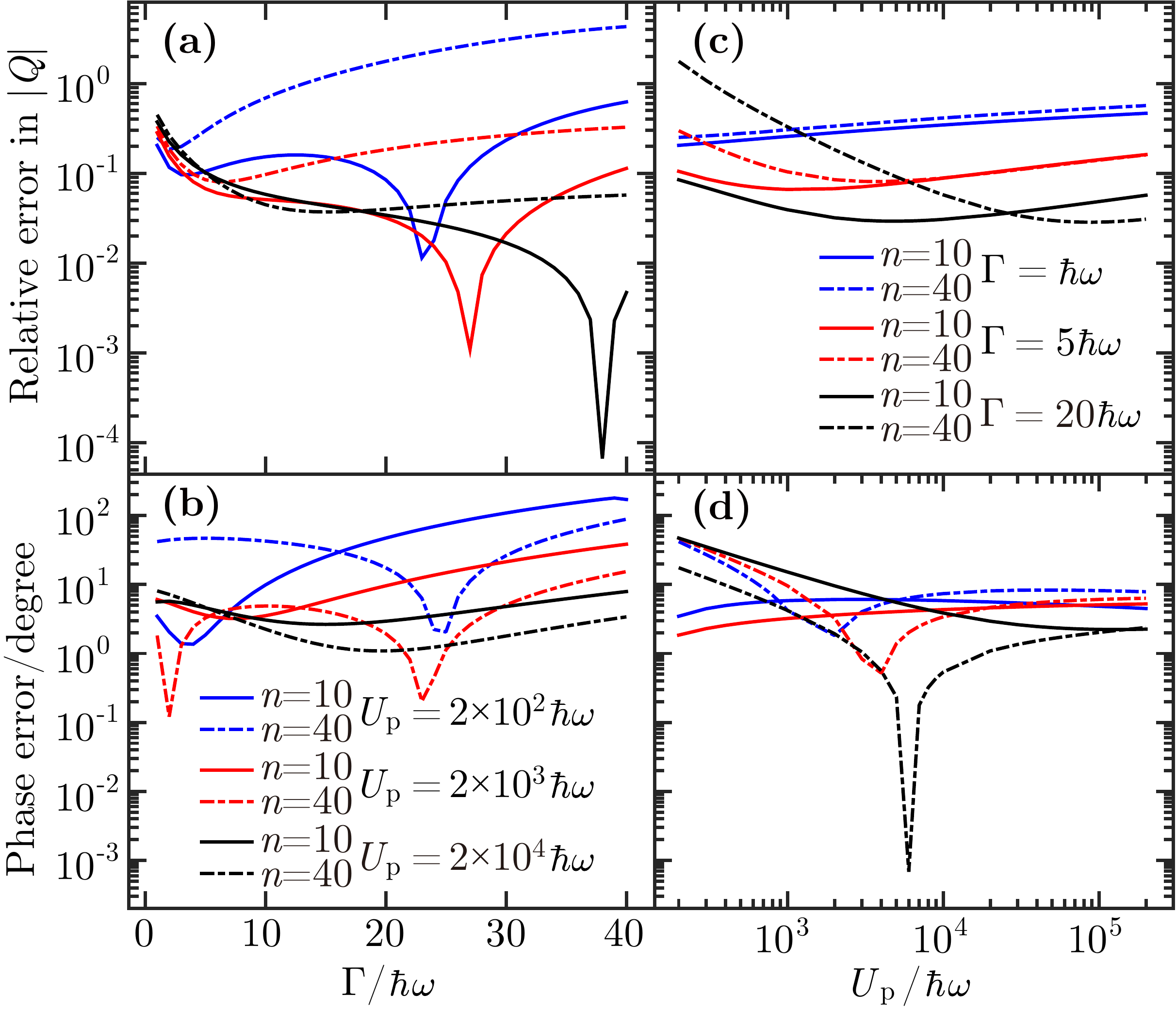}
	\caption{The accuracy of the linear-in-time approximation for the dimensionless sideband amplitude $Q_n$. (a) and (b) show respectively the relative errors in $|Q_n|$ and absolute errors in the phases as functions of the dephasing constant $\Gamma$ with ponderomotive energy $U_{\rm p}$ fixed at $2\times10^2\hbar\omega$ (blue curves), $2\times10^3\hbar\omega$ (red curves), and $2\times10^4\hbar\omega$ (black curves). (c) and (d) show respectively the relative errors in $|Q_n|$ and absolute errors in the phases as functions of the ponderomotive energy $U_{\rm p}$ with the dephasing constant $\Gamma$ fixed at $\hbar\omega$ (blue curves), $5\hbar\omega$ (red curves), and $20\hbar\omega$ (black curves). The results for sideband indices $n=10$ and $n=40$ are plotted as solid and dash-dotted curves, respectively. Zero detunings are used for all cases.}
	\label{FIG:linear_field_accuracy}
\end{figure}
To see whether the accuracy of the linear-in-time approximation remains high for a wide range of dephasing constants and ponderomotive energies, we compute the errors in the dimensionless sideband amplitudes $Q_n$ within the same parameter space as in the accuracy analysis of the saddle-point approximation shown in Fig.~\ref{FIG:saddle_point_accuracy}. Fig.~\ref{FIG:linear_field_accuracy} (a) and (b) show respectively the relative errors in the absolute values of $Q_n$ and the absolute errors in the phases of $Q_n$ as functions of the dephasing constant $\Gamma$ with the ponderomotive energy $U_{\rm p}$ fixed at $2\times10^2\hbar\omega$ (blue curves), $2\times10^3\hbar\omega$ (red curves), and $2\times10^4\hbar\omega$ (black curves). For the cases with the smallest ponderomotive energy, $U_{\rm p}=2\times10^2\hbar\omega$, the relative errors in $|Q_n|$ mostly stay above 10\% (blue curves in Fig.~\ref{FIG:linear_field_accuracy} (a)), and the phase errors can go up to around 200 degrees (blue curves in Fig.~\ref{FIG:linear_field_accuracy} (b)). For the cases with $U_{\rm p}=2\times10^3\hbar\omega$, the relative errors in $|Q_n|$ are also mostly above 10\% for the 40th-order sideband (red dash-dotted curve in Fig.~\ref{FIG:linear_field_accuracy} (a)), and the phase errors can get to about 40 degrees for the 10th-order sideband (red solid curve in Fig.~\ref{FIG:linear_field_accuracy} (b)). In contrast to the results in Fig.~\ref{FIG:saddle_point_accuracy} (a) and (b), which concern the accuracy of the saddle-point approximation, large ponderomotive energy is favored to achieve high accuracy in the linear-in-time approximation. Fig.~\ref{FIG:linear_field_accuracy} (c) and (d) show respectively the relative errors in $|Q_n|$ and the absolute errors in the phases of $Q_n$ as functions of the ponderomotive energy $U_{\rm p}$ with the dephasing constant $\Gamma$ fixed at $\hbar\omega$ (blue curves), $5\hbar\omega$ (red curves), and $20\hbar\omega$ (black curves). In the limit of large ponderomotive energy, both the relative errors in $|Q_n|$ and the phase errors match the results in Fig.~\ref{FIG:saddle_point_accuracy} (c) and (d). As the ponderomotive energy gets smaller, the accuracy of the linear-in-time approximation for the cases with relatively high sideband indices and strong dephasing gradually become lower than the limits set by the saddle-point approximation. Several dips corresponding to sign changes in the differences are also seen in Fig.~\ref{FIG:linear_field_accuracy} (a), (b), and (d).

\begin{figure}
	\includegraphics[width=0.47\textwidth]{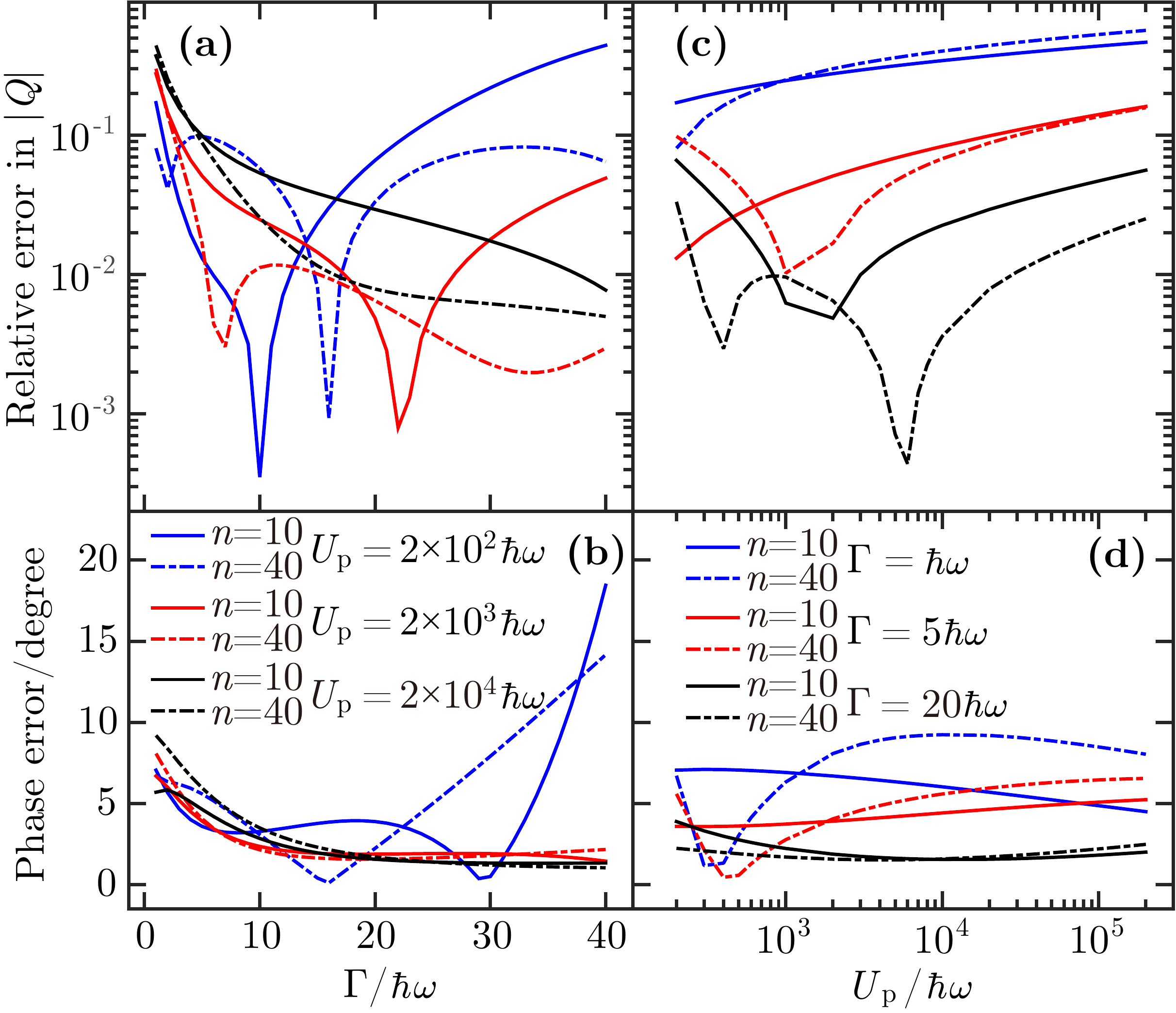}
	\caption{The accuracy of the linear-in-time approximation with a higher-order correction for the dimensionless sideband amplitude $Q_n$. (a) and (b) show respectively the relative errors in $|Q_n|$ and absolute errors in the phases as functions of the dephasing constant $\Gamma$ with ponderomotive energy $U_{\rm p}$ fixed at $2\times10^2\hbar\omega$ (blue curves), $2\times10^3\hbar\omega$ (red curves), and $2\times10^4\hbar\omega$ (black curves). (c) and (d) show respectively the relative errors in $|Q_n|$ and absolute errors in the phases as functions of the ponderomotive energy $U_{\rm p}$ with the dephasing constant $\Gamma$ fixed at $\hbar\omega$ (blue curves), $5\hbar\omega$ (red curves), and $20\hbar\omega$ (black curves). The results for sideband indices $n=10$ and $n=40$ are plotted as solid and dash-dotted curves, respectively. Zero detunings are used for all cases.}
	\label{FIG:linear_field_correction_accuracy}
\end{figure}

In order to obtain an algebraic form with higher accuracy, we introduce corrections up to the order of $(\hbar\omega/U_{\rm p})^{3/4}$ to the creation time $t'_n$, recombination time $t_n$, and the time duration $\tau_n$, which read (see Appendix~\ref{APP:correction_cubic} for the derivation)
\begin{align}
\omega \tilde{t}'_n
=
&
(\frac{2\hbar\omega}{9U_{\rm p}})^{1/4}
\frac{
2\zeta_0-\zeta_n
}
{
\sqrt
{
\zeta_n-\zeta_0
}
}
+(\frac{2\hbar\omega}{9U_{\rm p}})^{3/4}
\notag\\
&
\frac{
23\zeta^2_0
(
2\zeta_0-3\zeta_n
)
+\zeta^2_n
(30\zeta_0-17\zeta_n)
}
{120
(\zeta_n-\zeta_0)^{3/2}
},
\end{align}
\begin{align}
\omega \tilde{t}_n
=
&
(\frac{2\hbar\omega}{9U_{\rm p}})^{1/4}
\frac{
2\zeta_n-\zeta_0
}
{
\sqrt
{
\zeta_n-\zeta_0
}
}
-(\frac{2\hbar\omega}{9U_{\rm p}})^{3/4}
\notag\\
&
\frac{
\zeta^2_0
(
17\zeta_0-30\zeta_n
)
+23\zeta^2_n
(3\zeta_0-2\zeta_n)
}
{120
(\zeta_n-\zeta_0)^{3/2}
},
\end{align}
\begin{align}
\omega \tau_n
=
&
(\frac{18\hbar\omega}{U_{\rm p}})^{1/4}
{
\sqrt
{
\zeta_n-\zeta_0
}
}\notag\\
&
+(\frac{18\hbar\omega}{U_{\rm p}})^{3/4}
\frac{
7(\zeta^2_0+\zeta^2_n)-4\zeta_0\zeta_n
}
{360
\sqrt{\zeta_n-\zeta_0}
}.
\end{align}
Including a corresponding correction to the semiclassical action in $S^{(t_n,\tau_n)}(t_n,\tau_n)$, we arrive at a new algebraic form,
\begin{align}
\mathbb{P}_{n} 
\approx 
&
2i^n{\bf C}
\exp
\{
i[
q_{1/4}(n,\frac{i\Gamma+\Delta}{\hbar\omega})(\frac{\hbar\omega}{U_{\rm p}})^{1/4}
\notag\\
&
+
q_{3/4}(n,\frac{i\Gamma+\Delta}{\hbar\omega})(\frac{\hbar\omega}{U_{\rm p}})^{3/4}
]
\}
\notag\\
&
(\frac{U_{\rm p}}{\hbar\omega})^{\frac{D-2}{8}}
\frac{\exp[-i\arg[q_0(n,\frac{i\Gamma+\Delta}{\hbar\omega})]/2]}
{\sqrt{|q_0(n,\frac{i\Gamma+\Delta}{\hbar\omega})|}},
\label{EQ:new_algebraic_form}
\end{align}
which contains a new function,
\begin{align}
q_{3/4}(n,\frac{i\Gamma+\Delta}{\hbar\omega})
=
&
(\frac{1}{18})^{1/4}
\frac{1}{1260\sqrt{\zeta_n - \zeta_0}}[103(\zeta_n^2-\zeta_0^2)^2
\notag\\
&
+232\zeta_0\zeta_n(\zeta^2_0+\zeta^2_n)
-184\zeta_0^2\zeta^2_n].
\label{EQ:definition_q3o4}
\end{align}
In parallel with the accuracy analysis shown in Fig.~\ref{FIG:saddle_point_accuracy} and Fig.~\ref{FIG:linear_field_accuracy}, we compute the errors in the dimensionless sideband amplitudes $Q_n$ using the new algebraic form, Eq.~\ref{EQ:new_algebraic_form}. As shown in Fig.~\ref{FIG:linear_field_correction_accuracy} (a), the relative errors in the absolute values of $Q_n$ for the cases with relatively small dephasing are close to the limits set by the saddle-point approximation. For sufficiently strong dephasing, the relative errors in $|Q_n|$ stay below 10\% except for the case with sideband index $n=10$ and ponderomotive energy $U_{\rm p}=2\times10^2\hbar\omega$ (solid blue curve in Fig.~\ref{FIG:linear_field_correction_accuracy} (a)). The absolute errors in the phases of $Q_n$ are mostly less than 5 degrees for the three selected values of ponderomotive energy (Fig.~\ref{FIG:linear_field_correction_accuracy} (b)). Even for the cases with $U_{\rm p}=2\times10^2\hbar\omega$, the phase errors stay below 20 degrees (blue curves in Fig.~\ref{FIG:linear_field_correction_accuracy} (b)). As shown in Fig.~\ref{FIG:linear_field_correction_accuracy} (c) and (d), both the relative errors in $|Q_n|$ and the phase errors approach the results in Fig.~\ref{FIG:saddle_point_accuracy} (c) and (d) for a wide range of relatively large ponderomotive energies. The relative errors in $|Q_n|$ are mostly below 10\% for the selected cases with moderate dephasing (red and black curves in Fig.~\ref{FIG:saddle_point_accuracy} (c)), while the phase errors are less than 10 degrees for all three selected dephasing cases (Fig.~\ref{FIG:linear_field_correction_accuracy} (d)). This remarkable suppression of the errors by the correction term ends our derivation of the algebraic forms for the sideband polarization vectors. 

\section{Nonzero detunings}\label{SEC:nonzero_detuning}

To finalize our tailoring of the Feynman path integrals, we discuss the effects from nonzero detunings in this section. From the saddle-point equations, we see that the solution of the saddle points depends on the detuning through the kinetic energy $E_{\rm eh}[k_n(t'')]$ at the creation time $t'_n$ and recollision time $t_n$. An example of the semiclassical recollision pictures associated with a dephasing constant $\Gamma=5\hbar\omega$ and a negative detuning $\Delta=-2\hbar\omega$ is shown in Fig.~\ref{FIG:semiclassical_picture} (h) and (i) (dashed curves). The nonzero detuning further distorted the curves representing the complex electron-hole separation. As a new feature for the complex kinetic energy, the real part starts from the value of the detuning $\Delta$ and ends at the sideband offset energy subtracted by $\Delta$. 
For the derivation of the two algebraic forms, Eq.~\ref{EQ:algebraic_form} and~\ref{EQ:new_algebraic_form}, 
we have seen from previous discussions that the role of the detuning $\Delta$ has no essential difference from that of the dephasing constant $\Gamma$, since the sideband amplitudes depend on $\Gamma$ and $\Delta$ through analytic functions of the complex variable $i\Gamma+\Delta$. However, the question remains how the accuracy of the linear-in-time approximation depends on the detuning.

\begin{figure}
	\includegraphics[width=0.5\textwidth]{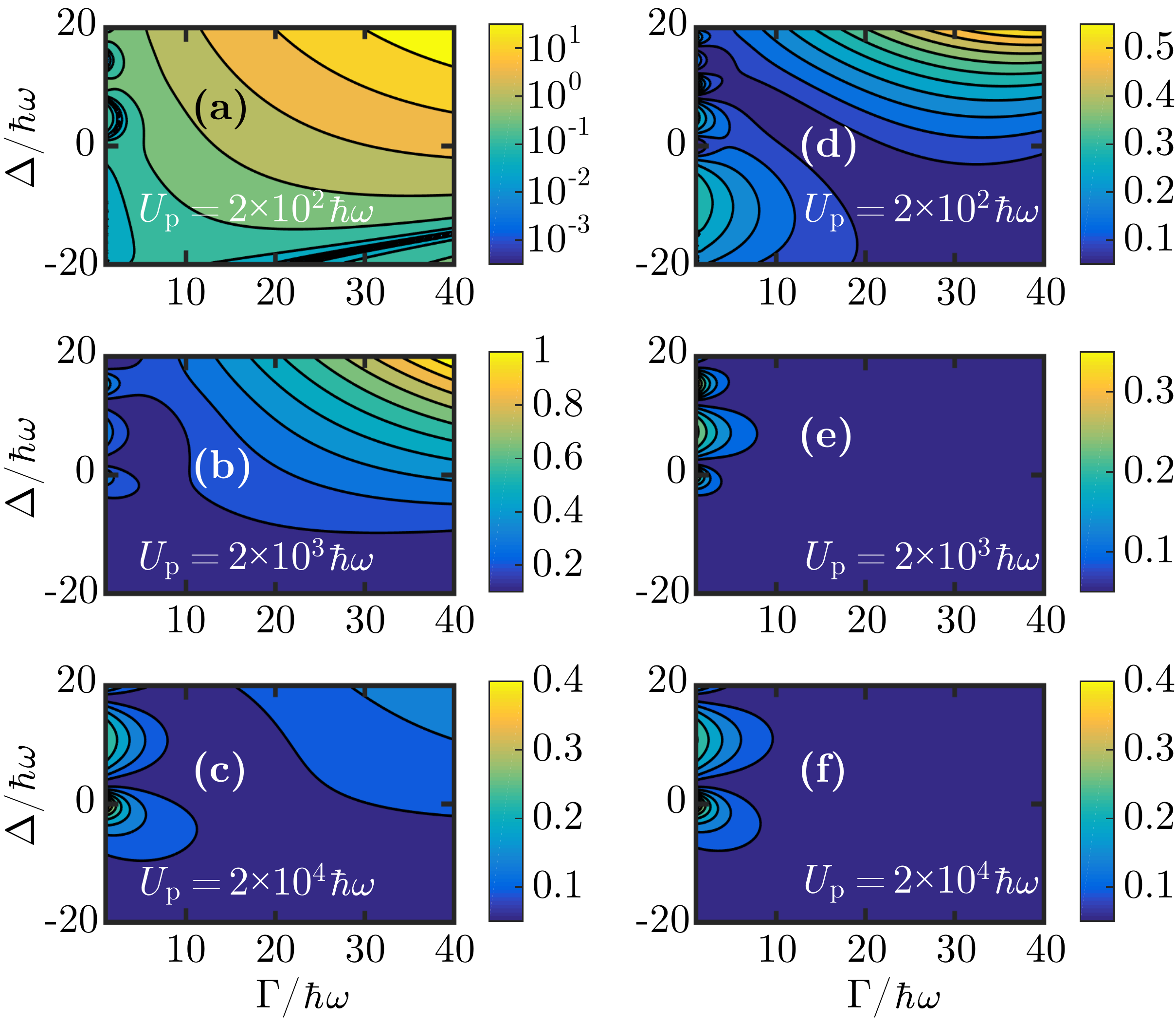}
	\caption{The accuracy of the linear-in-time approximation for the absolute values of the dimensionless sideband amplitudes $Q_{40}$ with varing dephasing and detuning. Left (Right) column: the relative errors in $|Q_{40}|$ without (with) a higher-order correction. The values of the ponderomotive energy $U_{\rm p}$ are chosen as $2\times10^2\hbar\omega$ ((a) and (d)), $2\times10^3\hbar\omega$ ((b) and (e)), and $2\times10^4\hbar\omega$ ((c) and (f)).}
	\label{FIG:detuning_abs}
\end{figure}
\begin{figure}
	\includegraphics[width=0.5\textwidth]{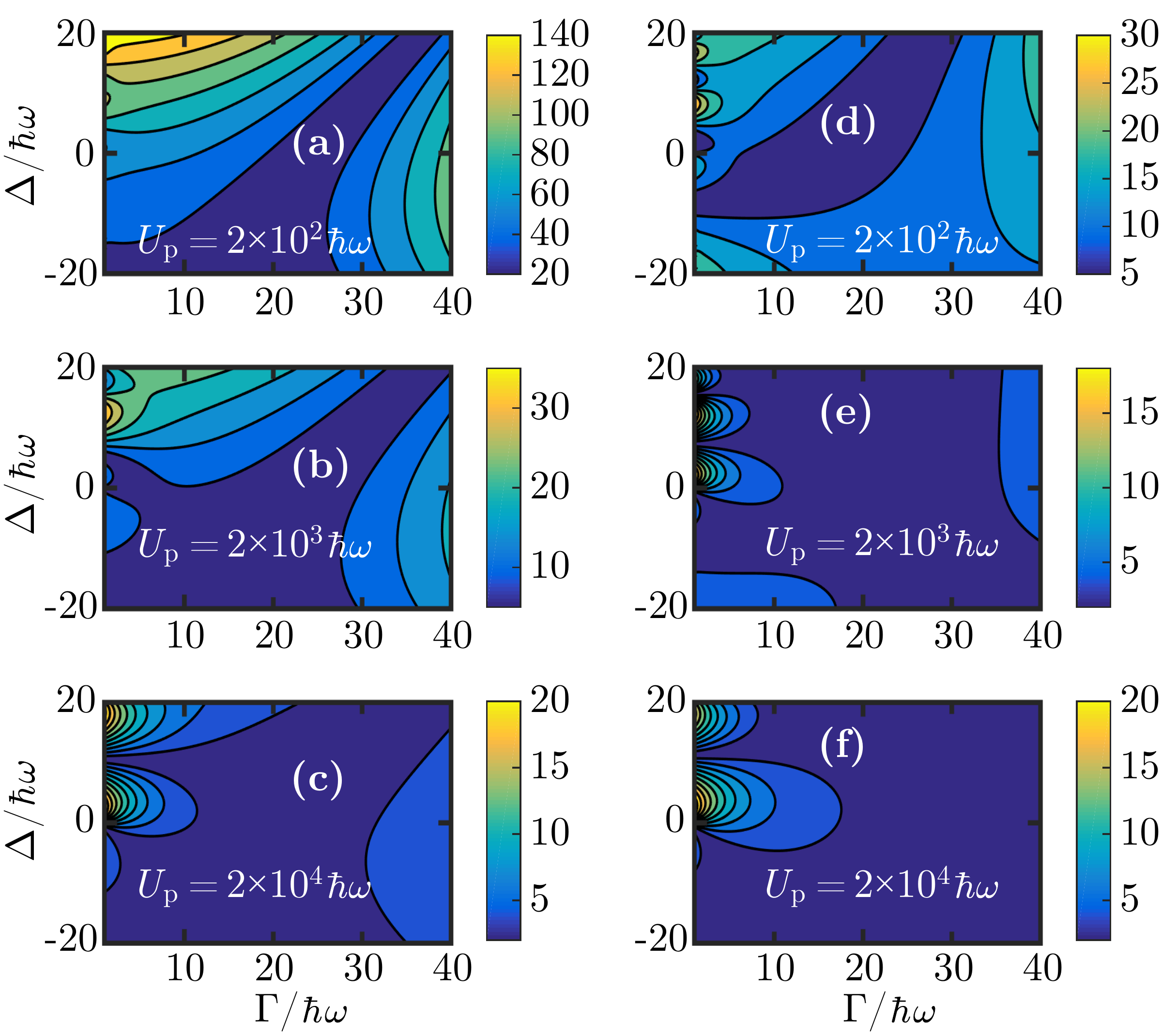}
	\caption{The accuracy of the linear-in-time approximation for the phases of the dimensionless sideband amplitudes $Q_{40}$ with varing dephasing and detuning. Left (Right) column: the absolute errors in the phases of $Q_{40}$ without (with) a higher-order correction. The values of the ponderomotive energy $U_{\rm p}$ are chosen as $2\times10^2\hbar\omega$ ((a) and (d)), $2\times10^3\hbar\omega$ ((b) and (e)), and $2\times10^4\hbar\omega$ ((c) and (f)).}
	\label{FIG:detuning_phase}
\end{figure}

To quantify the dependence of the accuracy of the linear-in-time approximation on the detuning, we compute the errors in the dimensionless sideband amplitudes $Q_n$ as functions of the dephasing constant $\Gamma\in[1,40]\hbar\omega$ and the detuning $\Delta\in[-20,20]\hbar\omega$, with the ponderomotive energy $U_{\rm p}$ fixed at three representative values, $2\times10^2\hbar\omega$, $2\times10^3\hbar\omega$, and $2\times10^4\hbar\omega$. Fig.~\ref{FIG:detuning_abs} and~\ref{FIG:detuning_phase} show respectively the relative errors in the absolute values of $Q_n$ and the absolute errors in the phases of $Q_n$ for sideband index $n=40$ (the results for $n=20,30$ are shown in Fig.~\ref{FIG:detuning_abs_n20}-\ref{FIG:detuning_phase_n30} in Appendix~\ref{APP:supplement_figures}). In each of the two figures, the errors in $Q_n$ calculated by using Eq.~\ref{EQ:algebraic_form} (Eq.~\ref{EQ:new_algebraic_form}) are presented in the left (right) column.
As shown in Fig.~\ref{FIG:detuning_abs} (a), for the cases with $U_{\rm p}=2\times10^2\hbar\omega$, the relative errors in $|Q_n|$ calculated by using Eq.~\ref{EQ:algebraic_form} are greater than 50\% in more than half of the parameter space investigated. As the ponderomotive energy increases to $2\times10^3\hbar\omega$, the relative errors in $|Q_n|$ are mostly less than 20\% (Fig.~\ref{FIG:detuning_abs} (b)). For the cases with $U_{\rm p}=2\times10^4\hbar\omega$, the relative errors in $|Q_n|$ stay below 10\% and can go even below 5\% in most of the parameter space (Fig.~\ref{FIG:detuning_abs} (c)). The correction term in Eq.~\ref{EQ:new_algebraic_form} greatly suppresses the relative errors in $|Q_n|$, as shown in Fig.~\ref{FIG:detuning_abs} (d), (e), and (f). The relative errors in $|Q_n|$ calculated by using Eq.~\ref{EQ:new_algebraic_form} can already go below 5\% in a wide range of dephasing constants and detunings for the cases with $U_{\rm p}=2\times10^2\hbar\omega$ (Fig.~\ref{FIG:detuning_abs} (d)). For the cases with the other two selected larger ponderomotive energies, the relative errors in $|Q_n|$ stay below 5\% in almost the whole parameter space (Fig.~\ref{FIG:detuning_abs} (e) and (f)). As shown in Fig.~\ref{FIG:detuning_phase}, the suppression of the phase errors in $|Q_n|$ by the correction term in Eq.~\ref{EQ:new_algebraic_form} is also remarkable. For the cases with $U_{\rm p}=2\times10^2\hbar\omega$, the phase errors calculated by using Eq.~\ref{EQ:algebraic_form} range from below 20 degrees to as large as 140 degrees in the parameter space investigated (Fig.~\ref{FIG:detuning_phase} (a)). For the cases with $U_{\rm p}=2\times10^3\hbar\omega$, the phase errors are mostly below 10 degrees (Fig.~\ref{FIG:detuning_phase} (b)). As the ponderomotive energy increases to $2\times10^4\hbar\omega$, the phase errors are mostly less than 5 degrees (Fig.~\ref{FIG:detuning_phase} (c)). In contrast, the phase errors calculated by using Eq.~\ref{EQ:new_algebraic_form} stay below 15 degrees in almost the whole parameter space shown in Fig.~\ref{FIG:detuning_phase} (d) for the cases with $U_{\rm p}=2\times10^2\hbar\omega$. For the cases with the other two selected larger ponderomotive energies, the phase errors are mostly less than 2.5 degrees, as shown in Fig.~\ref{FIG:detuning_phase} (e) and (f). The results are similar for two- and three-dimensional cases ($D=2,3$) (see Fig.~\ref{FIG:detuning_abs_n40_dimk2}-\ref{FIG:detuning_phase_n40_dimk3} in Appendix~\ref{APP:supplement_figures}). We thus see that the algebraic form, Eq.~\ref{EQ:new_algebraic_form}, is suitable for describing relatively low orders of sidebands in a wide range of parameters that are experimentally accessible.

\section{Feynman-path interferometer}\label{SEC:feynman_interferometer}

A straightforward application of our algebraic forms is to guide the control of the sideband amplitudes. The pump NIR laser does not need to be monochromatic. For instance, one can build up an interferometer using a NIR laser field with two central frequencies separated by an even number times of the THz frequency $\omega$, ${\bf E}_{\rm NIR}(t')={\bf F}_{{\rm NIR}}[1+\rho_{21}e^{-i(2N\omega t'-\varphi_{21})}]e^{-i\Omega t'}$, where $N$ is an integer, and the real parameters $\rho_{21}$ and $\varphi_{21}$ control respectively the relative strength and phase delay between the two frequency components. Two sets of sidebands produced respectively by the two frequency components of the NIR laser are located at the same frequencies, and thus interference occurs at each of the sideband frequencies. On the condition that the linear-in-time approximation is valid, as discussed earlier, for a monochromatic NIR laser, a shortest electron-hole recollision pathway dominantly contributes to each sideband amplitude within half a period of the THz field. Therefore, this interference can also be considered as the interference between two electron-hole recollision pathways. By using the algebraic form, Eq.~\ref{EQ:new_algebraic_form}, the resulting sideband polarization vector at frequency $\Omega+n\omega$ ($n$ is an even integer) can be written as
\begin{align}
\mathbb{P}(\Omega+n\omega)
&
\approx 
{\bf C}[
Q_n(\frac{i\Gamma+\Delta}{\hbar\omega},\frac{U_{\rm p}}{\hbar\omega})\notag\\
&
+\rho_{21} e^{i\varphi_{21}}
Q_{n-2N}(\frac{i\Gamma+\Delta}{\hbar\omega}+2N,\frac{U_{\rm p}}{\hbar\omega})],
\end{align}
which contains the detuning $\Delta=\hbar\Omega-E_{\rm g}$, and the dimensionless sideband amplitude in the form,
\begin{align}
Q_n(\frac{i\Gamma+\Delta}{\hbar\omega},& \frac{U_{\rm p}}{\hbar\omega})
=
2i^n
\exp
\{
i[
q_{1/4}(n,\frac{i\Gamma+\Delta}{\hbar\omega})(\frac{\hbar\omega}{U_{\rm p}})^{1/4}
\notag\\
&
+
q_{3/4}(n,\frac{i\Gamma+\Delta}{\hbar\omega})(\frac{\hbar\omega}{U_{\rm p}})^{3/4}
]
\}
\notag\\
&
(\frac{U_{\rm p}}{\hbar\omega})^{\frac{D-2}{8}}
\frac{\exp[-i\arg[q_0(n,\frac{i\Gamma+\Delta}{\hbar\omega})]/2]}
{\sqrt{|q_0(n,\frac{i\Gamma+\Delta}{\hbar\omega})|}}.
\end{align}
By varying the phase delay $\varphi_{21}$, the intensity of the sideband can be tuned between the values
\begin{align}
I_{n,\pm}
=
I_{n,0}
[1\pm
\rho_{21}
\frac{|Q_{n-2N}(\frac{i\Gamma+\Delta}{\hbar\omega}+2N,\frac{U_{\rm p}}{\hbar\omega})|}
{|Q_n(\frac{i\Gamma+\Delta}{\hbar\omega},\frac{U_{\rm p}}{\hbar\omega})|}]^2,
\label{EQ:interferometer_abs}
\end{align}
where $I_{n,0}$ is the sideband intensity when the second frequency component is switched off ($\rho_{21}=0$). The maximal sideband intensity is obtained when the two recollision pathways are in phase such that
\begin{align}
&\arg[Q_{n-2N}(\frac{i\Gamma+\Delta}{\hbar\omega}+2N, \frac{U_{\rm p}}{\hbar\omega})]
+\varphi_{21}\notag\\
=
&\arg[Q_n(\frac{i\Gamma+\Delta}{\hbar\omega}, \frac{U_{\rm p}}{\hbar\omega})]
\, ({\,\rm mod\,} 2\pi).
\label{EQ:interferometer_phase}
\end{align}

Such an interferometer can be used to extract the dephasing constant $\Gamma$, the bandgap $E_{\rm g}$ in the detuning $\Delta$ and the reduced mass $\mu$ in the ponderomotive energy $U_{\rm p}$. By measuring the maximal and minimal relative sideband intensities $I_{n,\pm}/I_{n,0}$ and the corresponding phase delays $\varphi_{21}$, two algebraic relations between the parameters $i\Gamma+\Delta$ and $U_{\rm p}$ can be seen from Eq.~\ref{EQ:interferometer_abs} and~\ref{EQ:interferometer_phase}. 
To determine the three real parameters, $\Gamma$, $\Delta$, and $U_{\rm p}$, it requires at least one additional equation, which can be obtained by adding a third frequency component of the NIR laser field. Although the absolute sideband intensity $I_{n,0}$ also contains information on the parameters $i\Gamma+\Delta$ and $U_{\rm p}$, determination of $I_{n,0}$ involves additional complexities such as modeling of the propagation of the NIR laser and sideband fields through optical setups. The absolute sideband intensity might also include a significant enhancement factor from electron-hole Coulomb interaction~\cite{yan2008theory}, which is outside the scope of this paper.

\section{Extracting material parameters by varying the TH\lowercase{z} field strength}\label{SEC:extract_parameters}

The dependence of the sideband intensities on the THz field strength~\cite{valovcin2018optical} provides a simpler way of extracting the dephasing constant $\Gamma$ and the reduced mass $\mu$ with a monochromatic NIR laser field. In cases where the algebraic form, Eq.~\ref{EQ:new_algebraic_form}, is valid, measuring intensities $I^{F_1}_{n}$ and $I^{F_2}_{n}$ of the $n$th-order sideband respectively for two THz field strengths $F_{{\rm max},1}$ and $F_{{\rm max},2}=\lambda F_{{\rm max},1}$ yields an algebraic equation for the parameters $i\Gamma+\Delta$ and $U_{\rm p}$,
\begin{align}
\sqrt{\frac{I^{F_2}_{n}}{I^{F_1}_{n}}}
=
\frac{|Q^{F_2}_{n}|}{|Q^{F_1}_{n}|}
=
&
\lambda^{\frac{D-2}{4}}
\exp
[
(1-\lambda^{-\frac{1}{2}})x_{1/4}
\notag\\
&
+
(1-\lambda^{-\frac{3}{2}})x_{3/4}
],
\end{align}
where we denote $Q^{F_s}_n\equiv Q_n({(i\Gamma+\Delta)}/{(\hbar\omega)},{U^{F_s}_{\rm p}}/{(\hbar\omega)})$ (s=1,2) and $x_{l}\equiv{\rm Im}[q_{l}(n,{(i\Gamma+\Delta)}/{(\hbar\omega)})]({\hbar\omega}/{U^{F_1}_{\rm p}})^{l}$ ($l=1/4,3/4$) with $U^{F_s}_{{\rm p}}\equiv{e^2F_{{\rm max},s}^2}/{(4\mu\omega^2)}$ being the ponderomotive energy corresponding to the THz field strength $F_{{\rm max},s}$. Taking the logarithm on both sides of the equation, we obtain an equation linear in the variables $x_{1/4}$ and $x_{3/4}$,
\begin{align}
&(1-\lambda^{-\frac{1}{2}})x_{1/4}
+
(1-\lambda^{-\frac{3}{2}})x_{3/4}
\notag\\
=
&
\frac{1}{2}\ln\frac{I^{F_2}_{n}}{I^{F_1}_{n}}-\frac{D-2}{4}\ln\lambda.
\label{EQ:intensity_fthz12}
\end{align}
Measuring the sideband intensities for three different THz field strengths produces two such equations, which can be easily solved for $x_{1/4}$ and $x_{3/4}$. The reduced mass can then be calculated as
\begin{align}
\mu=
\frac{e^2F_{{\rm max},1}^2}{4\hbar\omega^3}
\frac
{x^4_{1/4}}
{\{{\rm Im}[q_{1/4}(n,\frac{i\Gamma+\Delta}{\hbar\omega})]\}^4},
\label{EQ:fthz_mass}
\end{align}
where the parameter $i\Gamma+\Delta$ satisfies the algebraic equation
\begin{align}
\frac
{x^3_{1/4}}
{x_{3/4}}
=
\frac{\{{\rm Im}[q_{1/4}(n,\frac{i\Gamma+\Delta}{\hbar\omega})]\}^3}
{{\rm Im}[q_{3/4}(n,\frac{i\Gamma+\Delta}{\hbar\omega})]}.
\label{EQ:fthz_dephasing}
\end{align}
If the detuning $\Delta$ is known, one can easily extract the dephasing constant $\Gamma$ from Eq.~\ref{EQ:fthz_dephasing} and then calculate the reduced mass $\mu$ using Eq.~\ref{EQ:fthz_mass}.
The whole extraction procedure can still be applied even if the dephasing constant $\Gamma$ depends on the sideband index $n$. The applicability of the procedure relies on the premise that the theory agrees with experiments. Depending on the complexities in real experiments, modifications of our theory might be necessary. For example, in the presence of multiple dephasing mechanisms, a theory with a dephasing constant might not be able to explain the experimentally observed fall-offs of sideband intensities~\cite{banks2013terahertz}. A possible modification is to replace the dephasing factor $\Gamma\tau$ in the action $S_n({\bf P},t,\tau)$ in Eq.~\ref{EQ:action_original} by an integral $\int_{t-\tau}^t dt''\Gamma[{\bf k}(t'')]$ with $\Gamma$ becoming a function of the kinetic momentum ${\bf k}$. Whether the saddle-point analysis in this paper still applies after such a modification is an interesting question to be explored in future works.

For a multi-band system with more than one species of electron-hole pairs, interference of recollision pathways associated with different species of electron-hole pairs might provide extra equations to extract the bandgap $E_{\rm g}$. Such interference can be investigated systematically through the dynamical Jones matrices~\cite{banks2017dynamical}, each of which maps the electric field of the NIR laser into a sideband polarization vector. In the basis of circular polarizations, $\sigma_{\pm}$ with helicity $\pm1$ ($\sigma_{\pm}=\pm(\hat{x}\pm i\hat{y})/\sqrt{2}$ for light fields propagating along the z-axis), we can reorganize Eq.~\ref{EQ:sideband_amplitude} into the form,
\begin{align}
    		\begin{pmatrix}
    			P_{+,n}^{\rm HSG}\\
    			P_{-,n}^{\rm HSG}
    		\end{pmatrix} =
        \mathcal{T}_n
    		\begin{pmatrix}
    			F_{+}^{\rm NIR}\\
    			F_{-}^{\rm NIR}
    		\end{pmatrix},
\end{align}
where $P_{\pm,n}^{\rm HSG}$ and $F_{\pm}^{\rm NIR}$ denote respectively the $\sigma_{\pm}$ components of the sideband polarization vector $\mathbb{P}_n$ and the vector ${\bf F}_{\rm NIR}$ in the electric field of the NIR laser, and the dynamical Jones matrix $\mathcal{T}_n$ is a two-by-two matrix. For a general constant dipole vector ${\bf d}=d_{+}\sigma_{+}+d_{-}\sigma_{-}$, The dynamical Jones matrix $\mathcal{T}_n$ can be written as
\begin{align}
    \mathcal{T}_n
       =
  \bar{C} \mu^{D/2} Q_n
    	\begin{pmatrix}
    		|d_{-}|^2 & d^{*}_{-}d_{+}\\
    		d_{-}d^{*}_{+} & |d_{+}|^2
    	\end{pmatrix},
\label{EQ:tmatrix_single}
\end{align}
which includes the dimensionless sideband amplitude $Q_n({(i\Gamma+\Delta)}/{(\hbar\omega)},{U_{\rm p}}/{(\hbar\omega)})$ and a constant
\begin{align}
\bar{C}=\frac{-1}{\hbar\omega}  e^{-i{\pi D}/{4}} (\frac{\omega}{2\pi\hbar})^{{D}/{2}}.
\end{align}
Due to time-reversal symmetry, each electron-hole pair is usually accompanied by another pair with a complex conjugate dipole vector. As a result, the dynamical Jones matrix in Eq.~\ref{EQ:tmatrix_single} is modified as
\begin{align}
 \mathcal{T}_n
       =
  \bar{C}\mu^{D/2}Q_n
    	\begin{pmatrix}
    		|{\bf d}|^2 & 2d^{*}_{-}d_{+}\\
    		2d_{-}d^{*}_{+} & |{\bf d}|^2
    	\end{pmatrix}.
\end{align}
The dynamical Jones matrix for a simplest extension, where two species of electron-hole pairs move independently in their respective bands, can then be written as
\begin{align}
 \mathcal{T}_n
       =
  \bar{C}\sum_{j=1,2}
  \mu_j^{D/2} Q^{(j)}_{n}
    	 \begin{pmatrix}
    		|{\bf d}_j|^2 & 2d^{*}_{j,-}d_{j,+}\\
    		2d_{j,-}d^{*}_{j,+} & |{\bf d}_j|^2
    	\end{pmatrix},
\end{align}
which explicitly show how the recollision pathways associated with the two species of electron-hole pairs interfere with each other. We have labeled the two species of electron-hole pairs by $j=1,2$, and denoted $Q^{(j)}_{n}\equiv Q_n({(i\Gamma_{j,n}+\Delta_j)}/{(\hbar\omega)},{U_{{\rm p},j}}/{(\hbar\omega)})$. Each species of electron-hole pair is assigned a reduced mass $\mu_j$, a dephasing constant $\Gamma_{j,n}$ depending on the sideband index $n$, a detuning $\Delta_j$, a ponderomotive energy $U_{p,j}\equiv{e^2F_{\rm max}^2}/{(4\mu_j\omega^2)}$, and a dipole vector ${\bf d}_j=d_{j,+}\sigma_{+}+d_{j,-}\sigma_{-}$. Recent development of sideband polarimetry has enabled the determination of each dynamical Jones matrix up to a constant factor~\cite{banks2017dynamical,costello2021reconstruction}. The first row of the dynamical Jones matrix $\mathcal{T}_n$ provides two linear equations with respect to the quantities
$\mu_j^{D/2}Q^{(j)}_{n}$ ($j=1,2$) associated with the two species of electron-hole pairs. The two linear equations have a unique solution if the dipole vectors ${\bf d}_j$ ($j=1,2$) satisfy the condition of linear independence,
\begin{align}
    	\frac{d^{*}_{1,-}d_{1,+}}
	{|{\bf d}_1|^2}	
	\ne
	\frac{d^{*}_{2,-}d_{2,+}}
	{|{\bf d}_2|^2}.
\label{EQ:dipole_constraint}
\end{align}
According to the discussion at the beginning of this section, with the absolute value of the quantity $\mu_j^{D/2}Q^{(j)}_{n}$ determined up to a constant factor for three different THz field strengths, the algebraic form, Eq.~\ref{EQ:new_algebraic_form}, can be used to determine the reduced mass $\mu_j$ and dephasing constant $\Gamma_{j,n}$ as functions of the detuning $\Delta_j$ ($j=1,2$). For a fixed THz field strength, taking the ratio $\mu_1^{D/2}Q^{(1)}_{n}/(\mu_2^{D/2}Q^{(2)}_{n})$ yields a complex equation for the parameters $i\Gamma_{j,n}+\Delta_j$ and $U_{{\rm p},j}$ ($j=1,2$),
\begin{align}
\frac{\mu_1^{D/2}Q^{(1)}_{n}}{\mu_2^{D/2}Q^{(2)}_{n}}
=
&
(\frac{\mu_{1}}{\mu_{2}})^{\frac{3D+2}{8}}
\sqrt{\frac{|q^{(2)}_0|}
{|q^{(1)}_0|}}
\exp\{i
\frac{\arg[q^{(2)}_0]-\arg[q^{(1)}_0]}{2}\}\notag\\
&
\exp
\{
i[
q^{(1)}_{1/4}(\frac{\hbar\omega}{U_{{\rm p},1}})^{1/4}
+
q^{(1)}_{3/4}(\frac{\hbar\omega}{U_{{\rm p},1}})^{3/4}
\notag\\
&
-q^{(2)}_{1/4}(\frac{\hbar\omega}{U_{{\rm p},2}})^{1/4}
-
q^{(2)}_{3/4}(\frac{\hbar\omega}{U_{{\rm p},2}})^{3/4}
]
\},
\label{EQ:fthz_detuning}
\end{align}
where we denote $q^{(j)}_l\equiv q_{l}(n,{(i\Gamma_{j,n}+\Delta_j)}/{(\hbar\omega)})$ with $j=1,2$ and $l=0,1/4,3/4$. By treating the reduced mass $\mu_j$ and dephasing constant $\Gamma_{j,n}$ as functions of the detuning $\Delta_j$ determined for each species of the electron-hole pairs, Eq.~\ref{EQ:fthz_detuning} represents an algebraic relation between the two detunings $\Delta_1$ and $\Delta_2$. With the ratio $\mu_1^{D/2}Q^{(1)}_{n}/(\mu_2^{D/2}Q^{(2)}_{n})$ for another THz field strength, we expect that the detunings and thus the bandgap $E_{\rm g}$ might be fully determined. We leave the question on the uniqueness of the solution from this procedure for future discussion.

\section{Discussion}\label{SEC:discussion}

\subsection{Connection with existing HSG experiments}

Experimental observation of high-order sideband generation (HSG) has been reported in two classes of materials. The first class includes bulk gallium arsenide (GaAs)~\cite{zaks2013high,costello2021reconstruction} and GaAs-based quantum wells (QWs)~\cite{zaks2012experimental,banks2013terahertz,banks2017dynamical,valovcin2018optical}. The second class includes bulk and monolayer tungsten diselenide ($\rm WSe_2$)~\cite{langer2016lightwave,langer2018lightwave,borsch2020super,freudenstein2022attosecond}. Our two-band model is appropriate for describing HSG in the direct-gap materials such as narrow GaAs QWs~\cite{banks2017dynamical} and monolayer $\rm WSe_2$~\cite{borsch2020super}, which have isolated parabolic bands near the bandgaps. The recent experiments of sideband polarimetry have also indicated that HSG in bulk GaAs can be approximated as resulting from the interference of two electron-hole species that move independently in the THz field when the NIR laser is near-resonant with the bandgap~\cite{costello2021reconstruction}. This means that our results can also be applied to describe HSG in bulk GaAs for the cases of near-resonant excitation by the NIR laser.

For the validity of our formula, the required large ponderomotive energy $U_{\rm p}/\hbar\omega$ (in units of the THz photon energy $\hbar\omega$) has already been achieved for both classes of materials. In a recent HSG experiment in bulk GaAs~\cite{costello2021reconstruction}, a THz field with a frequency $f=\omega/(2\pi)=0.447$ THz and a field strength $F_{\max}=70$ kV/cm is used, corresponding to values of $U_{\rm p}/\hbar\omega$ being around 2500 and 3900 respectively for the two species of electron-hole pairs associated with two species of holes. The reduced masses for the two species of electron-hole pairs are taken respectively to be in the ranges $[0.057,0.061]m_0$ and $[0.037,0.038]m_0$ in the $k_x$-$k_y$ plane, where $m_0$ is the electron rest mass~\cite{vurgaftman2001band}. In a report of HSG in monolayer $\rm WSe_2$~\cite{borsch2020super}, a THz field with a frequency $f$ as low as 27 THz and a field strength as high as 19 MV/cm is applied, corresponding to $U_{\rm p}/\hbar\omega=291$ if the reduced mass is chosen as $\mu=0.17m_0$~\cite{berkelbach2013theory}. Therefore, according to the discussion in Section~\ref{SEC:extract_parameters}, experiment conditions are ready for testing our method of extracting the dephasing constant and reduced mass in monolayer $\rm WSe_2$, and extracting the dephasing constants, the bandgap and reduced masses in bulk GaAs. 

We expect our method can be used to extract dephasing constants and reduced masses in various direct-gap semiconducting and insulating materials that have isolated parabolic bands near the bandgaps. For direct-gap multi-band systems such as bulk GaAs, where two species of electron-hole pairs can be created and move independently in their respective bands, the bandgaps can also be extracted through our approach if the dipole vectors associated with the two electron-hole species satisfy Eq.~\ref{EQ:dipole_constraint}.

\subsection{Hints for more complicated systems}

In a general multi-band system, different electron-hole species can couple with each other while they are accelerated by the linearly-polarized THz field. In the limit of negligible carrier occupations, the sideband polarization vectors can still be expressed as Feynman path integrals under the approximation of free electrons and holes~\cite{banks2017dynamical}. However, the coupling between different electron-hole species results in the presence of non-Abelian Berry curvatures, which makes the analysis of the Feynman path integrals with the saddle-point method very complicated~\cite{banks2017dynamical}. It is still not clear if HSG for such systems can be described by the saddle-point method quantitatively. If the saddle-point approximation still applies, for sufficiently strong dephasing and sufficiently small kinetic energy gain, we expect that the semiclassical trajectories dominantly contributing to the sideband emission should still happen around the nodes of the THz field in order to get effective overlap between the electron and hole wavepackets, at least along the direction of the THz field. If this is true, one might be able to use linear-in-time (LIT) approximation to greatly simplify the analysis and reveal simple laws from the intricate HSG in multi-band systems with non-Abelian Berry curvature.

\subsection{Connection with HHG}

Due to the similarity between HSG and the interband processes in high-harmonic generation (HHG), our results can also be useful in the analysis of HHG if the interband processes dominate. For the readers who are familiar with the semiconductor Bloch equations (SBEs)~\cite{lindberg1988effective} but not the integral form of sideband polarization vectors, Eq.~\ref{EQ:sideband_amplitude}, we would like to mention that Eq.~\ref{EQ:sideband_amplitude} 
results from a summation of the microscopic polarization $p_{{\bf k}(t)}$ in the SBEs followed by a Fourier transform,
\begin{align}
\mathbb{P}_{n} 
= 
\frac{1}{T_{\rm THz}}\int_{0}^{T_{\rm THz}} dt e^{i(\Omega+n\omega )t} 
 \int \frac{d^D{\bf P}}{(2\pi)^D}
{\bf d}^{*}
p_{{\bf k}(t)},
\end{align}
The microscopic polarization $p_{{\bf k}(t)}$ has the form,
\begin{align}
p_{{\bf k}(t)}
= & \frac{i}{\hbar}
 \int_{-\infty}^{t} dt' 
{\bf d}\cdot {\bf E}_{\rm NIR}(t')\notag\\
 & \exp
\{
-\frac{i}{\hbar}\int_{t'}^t dt'' (E_{\rm cv}[{\bf k}(t'')]-i\Gamma)
\},
\label{EQ:micro_pk}
\end{align}
which satisfies one of the SBEs in the limit of negligible carrier occupations,
\begin{align}
i\hbar\frac{d }{dt}p_{{\bf k}(t)}
&
=i\hbar\frac{\partial}{\partial t}p_{{\bf k}(t)}+i\hbar\dot{\bf k}(t)\cdot\frac{\partial}{\partial {\bf k}}p_{{\bf k}(t)}
\notag\\
& 
=(E_{\rm cv}[{\bf k}(t)]-i\Gamma)p_{{\bf k}(t)}
-
{\bf d}\cdot {\bf E}_{\rm NIR}(t),
\label{EQ:pk_sbe}
\end{align}
where the Coulomb interaction is ignored and the scattering effects are described phenomenologically by the dephasing constant $\Gamma$. In HSG, the kinetic momentum $\hbar{\bf k}$ satisfies the equation of motion, $\hbar\dot{\bf k}(t)=-e{\bf F}_{\rm THz}(t)$. By substituting the THz and NIR laser fields with a single laser field, Eq.~\ref{EQ:pk_sbe} can also be used to describe the interband HHG in cases where the limit of negligible carrier occupations and the approximation of free electrons and holes still apply. For such cases, the interband polarization vectors are of the form
\begin{align}
\mathbb{P}^{\rm HHG}_{n} 
&
= 
\frac{i}{\hbar}\frac{1}{T_0}\int_{0}^{T_0} dt e^{i(n+1)\omega_0t} 
 \int \frac{d^D{\bf P}}{(2\pi)^D}
 \int_{-\infty}^{t} dt'  {\bf d}^{*}
\notag\\
& \exp
\{
-\frac{i}{\hbar}\int_{t'}^t dt'' (E_{\rm cv}[{\bf k}(t'')]-i\Gamma)
\}
{\bf d}\cdot {\bf F}_{0}(t'),
\end{align}
where $n$ is an even integer, and $T_0=2\pi/\omega_0$ is the period of the driving laser field ${\bf F}_0$. For a driving field of the form ${\bf F}_0(t)=\hat{x}F_{\rm max}\cos(\omega_0 t)$, the interband polarization $\mathbb{P}^{\rm HHG}_{n}$ contains two terms corresponding to the sideband polarization vector $\mathbb{P}_{n}$ in Eq.~\ref{EQ:sideband_amplitude} with the substitutions, ${\bf F}_{\rm NIR}\rightarrow \hat{x}F_{\rm max}/2$, $\Omega\rightarrow\pm\omega_0$, $\omega\rightarrow\omega_0$, $n\rightarrow (n+1)\mp1$ on the right-hand side of the equation. Therefore, our algebraic formulae for the sideband polarization vector $\mathbb{P}_{n}$, Eq.~\ref{EQ:algebraic_form},~\ref{EQ:definition_q1o4},~\ref{EQ:definition_q0},~\ref{EQ:new_algebraic_form} and~\ref{EQ:definition_q3o4}, can be directly applied in the analysis of the interband HHG under the aforementioned assumptions.

\section{Conclusion}\label{SEC:conclusion}

In summary, we have introduced a linear-in-time approximation and derived an explicit formula for electron-hole recollisions in a prototypical two-band model by tailoring Feynman path integrals. Our formula connects the sideband amplitudes with the laser-field and material parameters in a highly nontrivial manner. Over a wide range of dephasing constant, detuning, and ponderomotive energy, we show that both the absolute values and phases of the sideband polarization vectors can be quantitatively described by our algebraic formula with high accuracy. We demonstrate a way to control the sideband amplitudes by building up a Feynman-path interferometer that can be used to extract the dephasing constant, the bandgap, and the reduced mass. We also propose a method of extracting the dephasing constant and the reduced mass by simple algebraic calculation with sideband intensities measured for three THz field strengths. For a multi-band system such as bulk GaAs near-resonantly excited by the NIR laser, we show the possibility of extracting the dephasing constants, the bandgap, and the reduced masses through algebraic calculations. We have also discussed how our approach can be useful for analyses of HSG in more complicated systems, as well as HHG when interband processes dominate.

\section*{Acknowledgment}

We thank J. B. Costello and S. D. O'Hara for stimulating discussions. This work is funded by NSF-DMR 2004995.

\appendix

\section{Saddle-point method}\label{APP:Saddle_point}

In this appendix, we illustrate the details of using saddle-point method to calculate the sideband polarization vectors from Eq.~\ref{EQ:Feynman_path_int}. We will discuss the case where there is only one saddle point associated with each sideband.

We first expand the action $S_n({\bf P},t,\tau)$ into a Taylor series up to the second order in the variables, ${\bf P}$, $t$, and $\tau$, around the saddle point $({\bf P}_n,t_n,\tau_n)$ for the $n$th-order sideband, $S_n\approx S_{\rm sc}(P_{n},t_n,\tau_n)+\delta^2 S_n/2$, with a semiclassical action
\begin{align}
S_{\rm sc}(P_{n},t_n,\tau_n)
=
& 
n\hbar\omega t_n
-\int_{t_n-\tau_n}^{t_n} dt''\frac{\hbar^2}{2\mu}[P_{n}+\frac{e}{\hbar}A(t'')]^2\notag\\
& +i(\Gamma-i\Delta)\tau_n,
\end{align}
and a second-order term,
\begin{align}
\delta^2 S_n
=
&
-\frac{\hbar^2\tau_n}{\mu}({\bf P}-P_x\hat{x})^2
\notag\\
&
+\frac{\partial^2 S_{\rm sc}}{\partial P^2_n} \delta P^2
+2\delta\tau\frac{\partial^2 S_{\rm sc}}{\partial\tau_n\partial P_n}
\delta P
+2\delta t
\frac{\partial^2 S_{\rm sc}}{\partial t_n\partial P_n}
\delta P\notag\\
&+\frac{\partial^2 S_{\rm sc}}{\partial t_n^2}\delta t^2
+2\delta\tau\frac{\partial^2 S_{\rm sc}}{\partial\tau_n\partial t_n}\delta t
+\frac{\partial^2 S_{\rm sc}}{\partial \tau_n^2}\delta \tau^2.
\end{align}
where $\delta P=P_x-P_n$, $\delta t=t-t_n$, and $\delta \tau=\tau-\tau_n$. Note that the momentum $\hbar{\bf P}_n$ is along the x-axis, as is obvious from the first saddle-point equation, Eq.~\ref{EQ:x_recollision}. Extending the limits of the integrals to infinities, we obtain the following Gaussian integrals,
\begin{align}
\mathbb{P}_n
\approx 
&
{\bf d}^{*}{\bf d}\cdot {\bf F}_{\rm NIR}\frac{i\omega}{\pi\hbar}
\exp[{\frac{i}{\hbar}S_{\rm sc}(P_n,t_n,\tau_n)}]
\int_{-\infty}^{+\infty} d\delta\tau
\notag\\
& 
\int_{-\infty}^{+\infty} d\delta t
\int _{-\infty}^{+\infty} \frac{d^D{\bf P}}{(2\pi)^D}
 \exp[{\frac{i}{2\hbar}\delta^2 S_n}].
\end{align}
To do the integrals, we first make the quadratic form $\delta^2 S$ diagonal. Introducing the variable
\begin{align}
\bar{P}
=
\delta P
-\frac{\partial f_{P_n}}{\partial t_n}\delta t
-\frac{\partial f_{P_n}}{\partial\tau_n}\delta\tau,
\end{align}
where $f_{P_n}(t_n,\tau_n)$ is the solution of $P_n$ from the saddle-point equation $\partial_{P_n} S_{\rm sc}(P_n,t_n,\tau_n)=0$, we can write the second-order term $\delta^2 S$ in the form
\begin{align}
\delta^2S_n
=
&
-\frac{\hbar^2\tau_n}{\mu}({\bf P}-P_x\hat{x})^2
+\frac{\partial^2 S_{\rm sc}}{\partial P^2_n} \bar{P}^2
+\frac{\partial^2 S^{(t,\tau)}_{\rm sc}}{\partial t_n^2}\delta t^2\notag\\
&
+2\delta \tau\frac{\partial^2 S^{(t,\tau)}_{\rm sc}}{\partial\tau_n\partial t_n}\delta t
+\frac{\partial^2 S^{(t,\tau)}_{\rm sc}}{\partial \tau_n^2}\delta\tau^2.
\end{align}
where $S^{(t,\tau)}_{\rm sc}(t_n,\tau_n)=S_{\rm sc}(f_{P_n}(t_n,\tau_n),t_n,\tau_n)$. Through a second change of variables, $\bar{t}=\delta t-{\partial_{\tau_n} f_{t_n}}\delta \tau$, with
${f_{t_n}}(\tau_n)$ being the solution of $t_n$ from ${\partial_{t_n} S^{(t,\tau)}_{\rm sc}(t_n,\tau_n)}=0$, we obtain
the diagonal form
\begin{align}
\delta^2S_n
=
&-\frac{\hbar^2\tau_n}{\mu}({\bf P}-P_x\hat{x})^2\notag\\
&
+\frac{\partial^2 S_{\rm sc}}{\partial P^2_n} \bar{P}^2
+\frac{\partial^2 S^{(t,\tau)}_{\rm sc}}{\partial t_n^2}\bar{t}^2
+\frac{\partial^2 S^{(\tau)}_{\rm sc}}{\partial \tau_n^2}\delta\tau^2,
\end{align}
where $S^{(\tau)}_{\rm sc}(\tau_n)=S^{(t,\tau)}_{\rm sc}(f_{t_n}(\tau_n),\tau_n)$. The Gaussian integrals converge if ${\partial^2_{P_n} S_{\rm sc}}=-\hbar^2\tau_n/\mu$, ${\partial^2_{t_n} S^{(t,\tau)}_{\rm sc}}$, and ${\partial^2_{\tau_n} S^{(\tau)}_{\rm sc}}$ are all nonzero and their imaginary parts are all non-negative. Under these conditions, carrying out the Gaussian integrals yields
\begin{align}
&{\bf P}_{n} 
\approx 
2{\bf C}
\exp[{\frac{i}{\hbar}S^{(t,\tau)}_{\rm sc}(t_n,\tau_n)}]
\notag\\
&\frac{
e^{
-(i/2)[
D\arg(\tau_n)
+\arg({\partial^2_{t_n} S^{(t,\tau)}_{\rm sc}})
+\arg({\partial^2_{\tau_n} S^{(\tau)}_{\rm sc}})
]
}
}
{
\sqrt
{
|
(\omega\tau_n)^D
[{\partial^2_{(\omega t_n)} S^{(t,\tau)}_{\rm sc}}/\hbar]
[{\partial^2_{(\omega \tau_n)} S^{(\tau)}_{\rm sc}}/
\hbar]
|
}
},
\end{align}
which includes a constant vector
\begin{align}
{\bf C}=\frac{-1}{\hbar\omega}  e^{-i{\pi D}/{4}} (\frac{\mu \omega}{2\pi\hbar})^{{D}/{2}} {\bf d}^{*}{\bf d}\cdot {\bf F}_{\rm NIR}.
\end{align}
We have eliminated $P_n$ in the action $S_{\rm sc}(P_n,t_n,\tau_n)$ using the solution of the saddle-point equation $\partial_{P_n} S_{\rm sc}(P_n,t_n,\tau_n)=0$, 
\begin{align}
P_n
=
f_{P_n}(t_n,\tau_n)
=
\frac{e}{\hbar\tau_n}\int_{t_n-\tau_n}^{t_n} dt''A(t'').
\label{EQ:solution_pn}
\end{align}
The explicit form of $S^{(t,\tau)}_{\rm sc}(t_n,\tau_n)$ reads
\begin{align}
&S^{(t,\tau)}_{\rm sc}(t_n,\tau_n)\notag\\
=
&
n\hbar\omega t_n+[i\Gamma+\Delta+U_{\rm p}(\gamma^2(\omega\tau_n)-1)]\tau_n\notag\\
&
+U_{\rm p}\tau_n\alpha(\omega\tau_n)\gamma(\omega\tau_n)\cos[\omega(\tau_n-2t_n)],
\end{align}
where we have introduced the functions $\alpha(x)=\cos(x/2)-\gamma(x)$ and $\gamma(x)=\beta(x)/(x/2)$ with $\beta(x)=\sin(x/2)$. 
The second saddle-point equation,
${\partial_{t_n} S^{(t,\tau)}_{\rm sc}(t_n,\tau_n)}=\partial_{t_n} S_{\rm sc}(P_n,t_n,\tau_n)=0$, gives an implicit form of the function $f_{t_n}(\tau_n)$,
\begin{align}
\sin[\omega(\tau_n-2f_{t_n})]
=
\frac
{n\hbar\omega}
{4U_{\rm p}\alpha(\omega\tau_n)\beta(\omega\tau_n)},
\label{EQ:ftn_implicit}
\end{align}
from which we can calculate the explicit forms of the derivatives ${\partial^2_{(\omega t_n)} S^{(t,\tau)}_{\rm sc}}/\hbar$ and
${\partial^2_{(\omega \tau_n)} S^{(\tau)}_{\rm sc}}/
\hbar$ as
\begin{align}
\frac{1}{\hbar}\frac{\partial^2 S^{(t,\tau)}_{\rm sc}}{\partial{(\omega t_n)}^2}
=
&
2n
\cot[\omega(\tau_n-2t_n)],
\\
\frac{1}{\hbar}\frac{\partial^2 S^{(\tau)}_{\rm sc}}{\partial{(\omega \tau_n)}^2}
=
&
\frac{n}{2}[\frac{\alpha^2(\omega\tau_n)+\beta^2(\omega\tau_n)}{\omega\tau_n\alpha(\omega\tau_n)\beta(\omega\tau_n)}+1]\cot[\omega(\tau_n-2t_n)]\notag\\
&
+\frac{n}{2}[\frac{\alpha^2(\omega\tau_n)-\beta^2(\omega\tau_n)}{2\alpha(\omega\tau_n)\beta(\omega\tau_n)}]^2\tan[\omega(2t_n-\tau_n)]\notag\\
&
+\frac{U_{\rm p}}{\hbar\omega}\frac{\alpha^2(\omega\tau_n)-\beta^2(\omega\tau_n)}{\omega\tau_n}.
\end{align}
To determine $t_n$ and $\tau_n$, one can use Eq.~\ref{EQ:ftn_implicit}, together with the third saddle-point equation ${\partial_{\tau_n} S^{(t,\tau)}_{\rm sc}(t_n,\tau_n)}=\partial_{\tau_n} S_{\rm sc}(P_n,t_n,\tau_n)=0$, which can be written as
\begin{align}
\cos[\omega(\tau_n-2t_n)]
=
\frac
{\alpha^2(\omega\tau_n)+\beta^2(\omega\tau_n)-\xi}
{\alpha^2(\omega\tau_n)-\beta^2(\omega\tau_n)},
\end{align}
where $\xi=[i\Gamma+\Delta+(n/2)\hbar\omega]/U_{\rm p}$.

\section{Analytic calculations}\label{APP:Analytic_calculations}

In this appendix, we perform analytic calculations to simplify the expression of the sideband polarization vectors, Eq.~\ref{EQ:sideband_amplitude}, into an integral over a single variable. 

We consider a general polarization state for the THz field with a vector potential
\begin{align}
{\bf A}(t)
=
-\Lambda\frac{F_{\rm max}}{\omega}
[
\cos\phi\sin(\omega t)\hat{x}
+\sin\phi\sin(\omega t+\varphi)\hat{y}
], \label{vector_arb}
\end{align}
where $\Lambda=\sqrt{{2}/{(1+\sqrt{\kappa})}}$ with $\kappa=\cos^2\varphi+\cos^2(2\phi)\sin^2\varphi$ and $\phi\in[0,\pi/2]$. Integrating out all canonical momentum components except for the one along the x- and y-axis, we write the sideband polarization vector in the form,
\begin{align}
\mathbb{P}_{n} 
= & {\bf C}
\frac{2\pi\hbar}{\mu}
\int_{0}^{T_{\rm THz}} \frac{dt}{T_{\rm THz}}e^{i(\Omega+n\omega )t} 
 \int \frac{dP_xdP_y}{(2\pi)^2}
\notag\\
& \int_0^{+\infty} \frac{d\tau }{(\omega\tau)^{(D-2)/2}}
\exp[{\frac{i}{\hbar}\mathbb{S}(P_x,P_y,t,\tau)}],
\end{align}
where the action $\mathbb{S}(P_x,P_y,t,\tau)$ is quadratic in both $P_x$ and $P_y$,
\begin{align}
\mathbb{S}
=
&
-\hbar\Omega t
-[\frac{\hbar^2 (P_x^2+P_y^2)}{2\mu}-i\Gamma-\Delta+\Lambda^2U_{\rm p}]\tau\notag\\
&
+\frac{2\Lambda\hbar eF_{\rm THz}}{\mu\omega^2}
\{P_x\cos\phi\sin\frac{\omega\tau}{2}\sin[\omega(\frac{\tau}{2}-t)]\notag\\
&
+P_y\sin\phi\sin\frac{\omega\tau}{2}\sin[\omega(\frac{\tau}{2}-t)-\varphi]\}
\notag\\
&
+\frac{\Lambda^2U_{\rm p}}{\omega}\{\cos^2\phi\sin(\omega\tau)\cos[\omega(\tau-2t)]\notag\\
&
+\sin^2\phi\sin(\omega\tau)\cos[\omega(\tau-2t)-2\varphi]\}.
\end{align}
Integrating out $P_x$ and $P_y$ gives
\begin{align}
\mathbb{P}_{n} 
= & {\bf C}
\frac{\omega}{i}
\int_{0}^{T_{\rm THz}} \frac{dt}{T_{\rm THz}}e^{i(\Omega+n\omega )t} 
\notag\\
&
\int_0^{+\infty} \frac{d\tau }{(\omega\tau)^{D/2}}
\exp[{\frac{i}{\hbar}\mathbb{S}^{(t,\tau)}(t,\tau)}],
\end{align}
where
\begin{align}
\mathbb{S}^{(t,\tau)}(t,\tau)
=&
\sqrt{\kappa}\Lambda^2U_{\rm p}\tau\gamma(\omega\tau)\alpha(\omega\tau)\cos[\omega(\tau-2t)-\varphi+\eta]\notag\\
&-\hbar\Omega t+\mathbb{S}^{(\tau)}(\omega\tau)\hbar\omega\tau,
\end{align}
with the functions $\alpha$ and $\gamma$ defined in Appendix.~\ref{APP:Saddle_point}, $\mathbb{S}^{(\tau)}(\omega\tau)\equiv (i\Gamma+\Delta)/(\hbar\omega)+\Lambda^2[U_{\rm p}/(\hbar\omega)][\gamma^2(\omega\tau)-1]$, and a constant $\eta$ defined by $\cos\eta={\cos\varphi}/{\sqrt{\kappa}}$ and $\sin\eta={\cos2\phi\sin\varphi}/{\sqrt{\kappa}}$.
Using the identity with the Bessel functions of the first kind, $J_m$,
\begin{align}
e^{iz\cos\theta}=\sum_{m=-\infty}^{+\infty}J_m(z)i^me^{im\theta},\label{exp_besselcos}
\end{align}
we arrive at a Fourier series,
\begin{align}
&\exp[{\frac{i}{\hbar}\mathbb{S}^{(t,\tau)}(t,\tau)}]\notag\\
=
&
\sum_{m}
e^{-i(\Omega+2m\omega) t}
i^m
J_m[\sqrt{\kappa}\Lambda^2\frac{U_{\rm p}}{\hbar}\tau\gamma(\omega\tau)\alpha(\omega\tau)]\notag\\
&
e^{im(\eta-\varphi)}
\exp\{i[\mathbb{S}^{(\tau)}(\omega\tau)+m]\omega\tau\},
\end{align}
from which we can immediately see that the sideband amplitudes are identically zero for odd sideband indices, while for even sideband indices, we obtain the following integral form,
\begin{align}
\mathbb{P}_{n} 
= & {\bf C}
i^{n/2-1}
\int_0^{+\infty} \frac{d(\omega\tau) }{(\omega\tau)^{D/2}}
J_{n/2}[\sqrt{\kappa}\Lambda^2\frac{U_{\rm p}}{\hbar\omega}\omega\tau\gamma(\omega\tau)\alpha(\omega\tau)]\notag\\
&
e^{i(n/2)(\eta-\varphi)}\exp\{i[\mathbb{S}^{(\tau)}(\omega\tau)+n/2]\omega\tau\}.
\label{EQ:integral_form_general}
\end{align}

For circularly polarized THz fields, we have $\phi={\pi}/{4}$ and $\varphi=\pm{\pi}/{2}$ so that $\kappa=0$, which implies that the sideband amplitudes are identically zero since the Bessel functions of nonzero integer orders satisfy $J_n(0)=0$.

For a linearly polarized THz field with vector potential ${\bf A}=-({F_{\rm max}}/{\omega})\cos(\omega t)\hat{x}$, we have $\eta=\varphi=0$ and $\kappa=\Lambda=1$ thus Eq.~\ref{EQ:integral_form_general} can be simplified as
\begin{align}
\mathbb{P}_{n} 
= & {\bf C}
i^{n/2-1}
\int_0^{+\infty} \frac{d(\omega\tau) }{(\omega\tau)^{D/2}}
J_{n/2}[\frac{U_{\rm p}}{\hbar\omega}\omega\tau\gamma(\omega\tau)\alpha(\omega\tau)]\notag\\
&
\exp\{i[\mathbb{S}^{(\tau)}(\omega\tau)+n/2]\omega\tau\},
\label{EQ:integral_form}
\end{align}
with $\mathbb{S}^{(\tau)}(\omega\tau)= (i\Gamma+\Delta)/(\hbar\omega)+[U_{\rm p}/(\hbar\omega)][\gamma^2(\omega\tau)-1]$.

\section{Maximum electron-hole separations and electron-hole wavefunction widths}\label{APP:excursion_vs_wavepacket}

In this appendix, we discuss the maximum electron-hole separations and electron-hole wavefunction widths for one-dimensional momentum space to gain some insights into how the accuracy of the saddle-point approximation depends on the dephasing constant $\Gamma$, the sideband index $n$, and the ponderomotive energy $U_p$. Intuitively, one expects that the recollision processes in HSG can be described by the semiclassical trajectories given by the saddle-point solutions if the maximum separations of the electron-hole pairs are much larger than the widths of their wavefunctions in real space.

We estimate the maximum electron-hole separations for the shortest classical recollision pathways within the linear-in-time approximation. Along a shortest classical recollision pathway, an electron and a hole are created with zero relative kinetic momentum ($\hbar k_n(t'_n)=0$), and the maximum separation is reached at $t_{\rm max}$ when the kinetic momentum $\hbar k_n(t)$ goes back to zero. Under the linear-in-time approximation, from Eq.~\ref{EQ:kn_t} and~\ref{Eq:t_0_linear} with $\Gamma=\Delta=0$, we see that
\begin{align}
\omega \tilde{t}'_n
=
-\omega  \tilde{t}_{\rm max}
=
-(\frac{2n\hbar\omega}{9U_{\rm p}})^{1/4}.
\end{align}
Integrating the relative velocity $v_{\rm eh}(t'')={\hbar k_n(t'')}/{\mu}$ from $t'_n$ to $t_{\rm max}$, we obtain the maximum electron-hole separation as
\begin{align}
x_{\rm max}
=
|\int_{t'_n}^{t_{\rm max}}dt''\frac{\hbar k(t'')}{\mu}|
=
\frac{2\sqrt{3}}{9P_{\rm max}}
(\frac{8n^3U_{\rm p}}{\hbar\omega})^{1/4},
\end{align}
where $\hbar P_{\rm max}=eF_{\rm max}/\omega$ is the maximum relative momentum obtainable from the THz field.

Next, we calculate the electron-hole wavefunction widths along the THz-field driving direction for one-dimensional momentum space. The electron-hole wavefunctions are equivalent to the microscopic polarization $p_{{\bf k}(t)}$ in Eq.~\ref{EQ:micro_pk}~\cite{banks2017dynamical}. For one-dimensional momentum space, the electron-hole wavefunctions can be calculated as
\begin{align}
p_{k(t)}
=
\frac{i}{\hbar}
{\bf d}\cdot {\bf F}_{\rm NIR}
\int_0^{+\infty}
d\tau 
e^{i\mathbb{S}(P,t,\tau)-i\Omega t},
\end{align}
with an action
\begin{align}
\mathbb{S}(P,t,\tau)
&
=
-[(\frac{2P^2}{P^2_{\rm max}}+1)U_{\rm p}-(i\Gamma+\Delta)]\frac{\tau}{\hbar}
\notag\\
&
+
\frac{ 
8U_{\rm p}}{\hbar\omega}
\frac{P}{P_{\rm max}}
\sin\frac{\omega\tau}{2}\sin[\omega(\frac{\tau}{2}-t)]
\notag\\
&
+
\frac{U_{\rm p}}{\hbar\omega}\sin(\omega\tau)
\cos[2\omega(\frac{\tau}{2}-t)].
\end{align}
Using the identity, Eq.~\ref{exp_besselcos}, we have the expansion,
\begin{align}
e^{i\mathbb{S}(P,t,\tau)}
=
&e^{-i[({2P^2}/{P^2_{\rm max}}+1)U_{\rm p}-(i\Gamma+\Delta)]({\tau}/{\hbar})}\notag\\
&\sum_{n_1}J_{n_1}[\frac{U_{\rm p}}{\hbar\omega}\sin(\omega\tau)]i^{n_1}e^{in_1\omega(\tau-2t)}\notag\\
&\sum_{n_2}J_{n_2}[\frac{8U_{\rm p}}{\hbar\omega}
\frac{P}{P_{\rm max}}\sin\frac{\omega\tau}{2}]
e^{in_2\omega(\frac{\tau}{2}-t)}.
\end{align}
Since the Bessel function $J_{n_2}(x)$ is even (odd) for even (odd) $n_2$, the terms with odd $n_2$ do not contribute to sideband generation because of inversion symmetry. Including only the terms with even $n_2$, we arrive at the following form of the electron-hole wavefunctions,
\begin{align}
p_{k(t)}
=
&
\frac{i}{\hbar\omega}
{\bf d}\cdot {\bf F}_{\rm NIR}
\sum_{n\,{\rm even}} 
\Psi_P(n)
e^{-i(\Omega+n\omega )t},
\end{align}
where each sideband frequency $\Omega+n\omega$ is associated with a momentum distribution function,
\begin{align}
\Psi_{P}(n)
=
&\int_0^{+\infty}
d(\omega\tau)e^{-i[2{P^2}/{P^2_{\rm max}}+1)U_{\rm p}-n{\hbar\omega}-(i\Gamma+\Delta)]({\tau}/{\hbar})}\notag\\
&\sum_{n'}
J_{2n'}[\frac{8U_{\rm p}}{\hbar\omega}
\frac{P}{P_{\rm max}}\sin\frac{\omega\tau}{2}]\notag\\
&J_{n-n'}[\frac{U_{\rm p}}{\hbar\omega}\sin(\omega\tau)]i^{n-n'}.
\end{align}
\begin{figure}
	\includegraphics[width=0.45\textwidth]{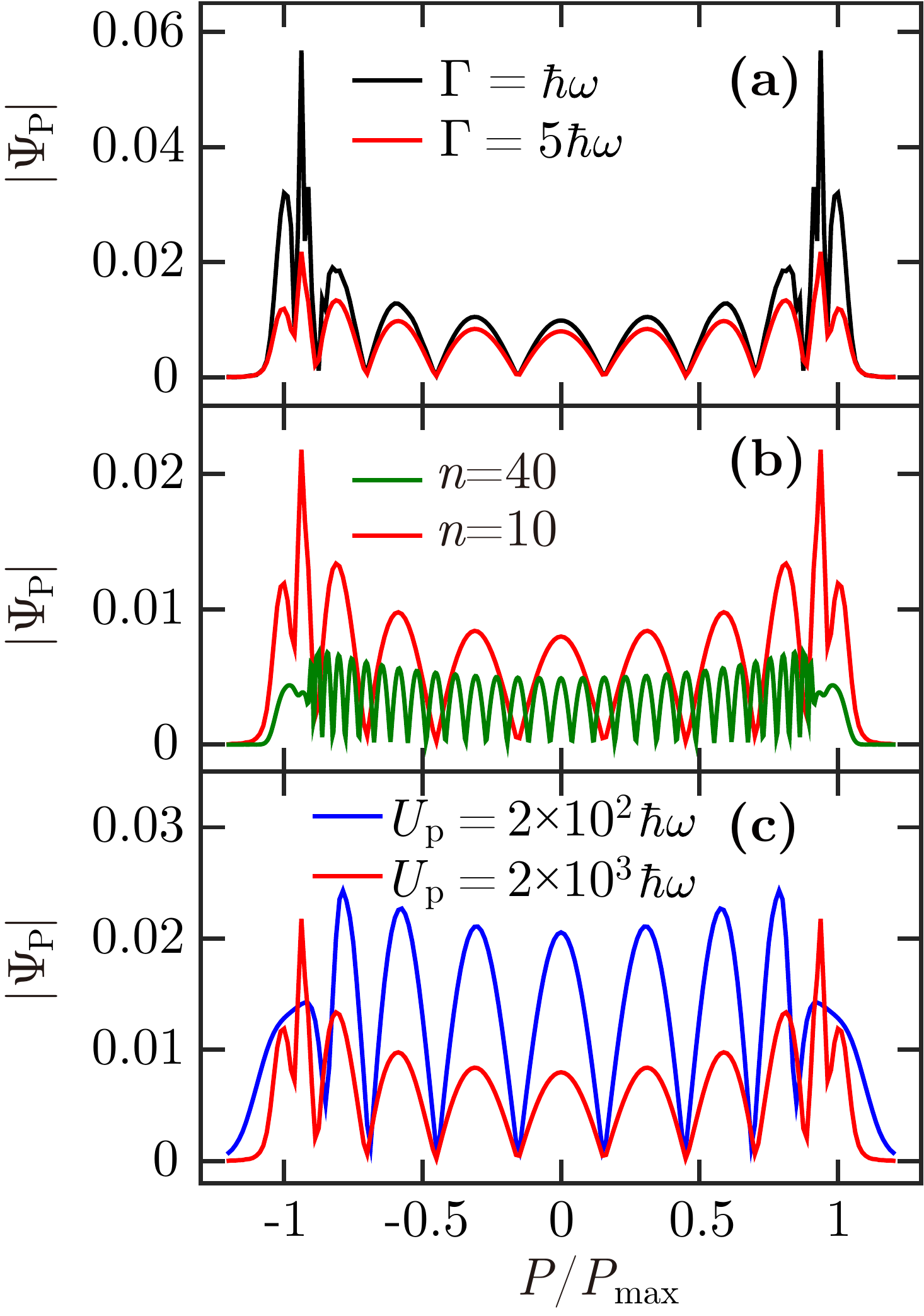}
	\caption{Momentum distributions of electron-hole wavefunctions. (a) The momentum distribution functions $\Psi_{P}(n)$ for two dephasing constants, $\Gamma=\hbar\omega$ (black curve) and $\Gamma=5\hbar\omega$ (red curve). (b) The momentum distribution functions $\Psi_{P}(n)$ for $n=10$ (red curve) and $n=40$ (dark green curve). (c) The momentum distribution functions $\Psi_{P}(n)$ for two values of the ponderomotive energy, $U_{\rm p}=2\times 10^2\hbar\omega$ (blue curve) and $U_{\rm p}=2\times 10^3\hbar\omega$ (red curve).  The red curves in (a), (b), and (c) represent the same momentum distribution function $\Psi_{P}(n)$ calculated for the 10th-order sideband with parameters $U_{\rm p}=2\times 10^3\hbar\omega$, $\Gamma=5\hbar\omega$, and $\Delta=0$. The two curves in each frame are calculated by using the same parameters except for the one shown in the legend.}
	\label{FIG:psi_p}
\end{figure}
Fig.~\ref{FIG:psi_p} (a), (b), and (c) show respectively the dependences of the momentum distribution function $\Psi_{P}(n)$ on the dephasing constant $\Gamma$, the sideband index $n$, and the ponderomotive energy $U_{\rm p}$. We observe that the momentum distribution function $\Psi_{P}(n)$ tends to be more localized for weaker dephasing, smaller sideband index, and larger ponderomotive energy. The peaks at around $\pm P_{\rm max}$ correspond to the saddle-point solution $P_n$ in Eq.~\ref{EQ:solution_pn} and its inverse.

Since the maximum separation $x_{\rm max}$ (in units of $P^{-1}_{\rm max}$) is larger for higher-order sidebands and larger ponderomotive energy, one expects that the saddle-point approximation should be of higher accuracy for relatively high-order sidebands and relatively strong dephasing, while the dependence of the accuracy on the ponderomotive energy relies on the competition between the maximum electron-hole separations and the electron-hole wavefunction widths.

\section{Corrections to the linear-in-time approximation}\label{APP:correction_cubic}

In this appendix, we derive the correction term to the linear-in-time approximation in the algebraic form, Eq.~\ref{EQ:new_algebraic_form}. 

The THz field strength near the node at $\omega t=\pi/2$ can in general be expanded in a Taylor series,
\begin{align}
F_{\rm THz}(t)=-F_{\rm max}[\omega \tilde{t}-\frac{(\omega \tilde{t})^3}{6}+\cdots].
\end{align}
From the Newtonian equation of motion $\hbar \dot{k}_n(t)=-eF_{\rm THz}(t)$, the kinetic momentum $\hbar {k}_n(t)$ can also be written as a Taylor series,
\begin{align}
\hbar k_n(t)=
&
\hbar k_n(t'_n)
+\frac{eF_{\rm max}}{\omega}
\{
\frac{1}{2}[(\omega \tilde{t})^2-(\omega  \tilde{t}'_n)^2]\notag\\
&
-\frac{1}{24}[(\omega \tilde{t})^4-(\omega  \tilde{t}'_n)^4+\cdots]
\}.
\end{align}
Putting this solution into the first saddle-point equation, Eq.~\ref{EQ:x_recollision}, yields
\begin{align}
\zeta_0\sqrt{\frac{\hbar\omega}{2U_{\rm p}}}
=
&
\frac{1}{6}[(\omega \tilde{t}_n)-(\omega  \tilde{t}'_n)]
\{
[(\omega \tilde{t}_n)+2(\omega  \tilde{t}'_n)]\notag\\
&
-\frac{1}{20}[(\omega \tilde{t}_n)^3+2(\omega \tilde{t}_n)^2(\omega \tilde{t}'_n)
+3(\omega \tilde{t}_n)(\omega \tilde{t}'_n)^2\notag\\
&
+4(\omega \tilde{t}'_n)^3]+\cdots
\}.
\label{EQ:general_t_n_1}
\end{align}
The solution of $k_n(t)$ at $t_n$ provides another equation for the time variables $\tilde{t}'_n$ and $\tilde{t}_n$,
\begin{align}
(\zeta_n+\zeta_0)\sqrt{\frac{\hbar\omega}{2U_{\rm p}}}
=
&
\frac{1}{2}[(\omega \tilde{t}_n)^2-(\omega  \tilde{t}'_n)^2]\notag\\
&
-\frac{1}{24}[(\omega \tilde{t}_n)^4-(\omega  \tilde{t}'_n)^4+\cdots].
\label{EQ:general_t_n_2}
\end{align}
Here we have used Eq.~\ref{EQ:k_n_0} and~\ref{EQ:k_n_c} to eliminate the kinetic momenta at $t'_n$ and $t_n$. To obtain the correction terms of higher-order in ${\hbar\omega}/{U_{\rm p}}$ to the solutions of $t'_n$ and $t_n$, we start a perturbation theory from the ansatzes, 
\begin{align}
\omega \tilde{t}'_n
&
=
\delta'_{1/4}
+
\delta'_{3/4},
\\
\omega \tilde{t}_n
&
=
\delta_{1/4}
+
\delta_{3/4},
\end{align}
where the factors $\delta'_{1/4}$ and $\delta_{1/4}$ are the solutions of $\omega \tilde{t}'_n$ and $\omega \tilde{t}_n$ of the order $({\hbar\omega}/{U_{\rm p}})^{1/4}$ under the linear-in-time approximation, given by Eq.~\ref{Eq:t_0_linear} and~\ref{Eq:t_c_linear}, and the factors $\delta'_{3/4}$ and $\delta_{3/4}$ are correction terms of the order $({\hbar\omega}/{U_{\rm p}})^{3/4}$.
Putting these ansatzes into Eq.~\ref{EQ:general_t_n_1} and~\ref{EQ:general_t_n_2} and keeping the lowest-order terms in ${\hbar\omega}/{U_{\rm p}}$, we obtain the following two linear equations with respect to the variables $\delta'_{3/4}$ and $\delta_{3/4}$,
\begin{align}
&
(2\delta_{1/4}
+\delta'_{1/4})\delta_{3/4}
+(\delta_{1/4}
-4\delta'_{1/4})\delta'_{3/4}\notag\\
=
&
\frac{1}{20}(\delta_{1/4}
-\delta'_{1/4})[\delta_{1/4}^3
+2\delta_{1/4}^2\delta'_{1/4}
\notag\\
&
+3\delta_{1/4}(\delta'_{1/4})^2+4(\delta'_{1/4})^3],
\end{align}
\begin{align}
\delta_{1/4}\delta_{3/4}-\delta'_{1/4}\delta'_{3/4}
=
\frac{1}{24}[\delta_{1/4}^4-(\delta'_{1/4})^4].
\end{align}
Solving these linear equations yields
\begin{align}
\delta_{3/4}
=
&
\frac{1}{120}
[
5\delta_{1/4}^3
-4\delta_{1/4}^2\delta'_{1/4}\notag\\
&
-7\delta_{1/4}(\delta'_{1/4})^2
-4(\delta'_{1/4})^3
],
\\
\delta'_{3/4}
=
&
\frac{1}{120}
[
5(\delta'_{1/4})^3
-4(\delta'_{1/4})^2\delta_{1/4}\notag\\
&
-7(\delta'_{1/4})\delta_{1/4}^2
-4(\delta_{1/4})^3
].
\end{align}
Substituting $\delta'_{1/4}$ and $\delta_{1/4}$ with the right-hand sides of Eq.~\ref{Eq:t_0_linear} and~\ref{Eq:t_c_linear}, after some straightforward algebra, we obtain
\begin{align}
\delta'_{3/4}
=
&
(\frac{2\hbar\omega}{9U_{\rm p}})^{3/4}
\frac{1}
{120(\zeta_n-\zeta_0)^{3/2}}\notag\\
&
[23\zeta^2_0
(
2\zeta_0
-3\zeta_n
)
+\zeta^2_n
(30\zeta_0
-17\zeta_n)]
,\\
\delta_{3/4}
=
&
-(\frac{2\hbar\omega}{9U_{\rm p}})^{3/4}
\frac{1}
{120(\zeta_n-\zeta_0)^{3/2}}\notag\\
&
[\zeta^2_0
(
17\zeta_0-30\zeta_n
)
+23\zeta^2_n
(3\zeta_0-2\zeta_n)].
\end{align}
To derive the correction term to the semiclassical action $S^{(t,\tau)}_{\rm sc}(t_n,\tau_n)$ of the order $({\hbar\omega}/{U_{\rm p}})^{3/4}$, we approximate the semiclassical action as the following Taylor polynomial,
\begin{align}
&\frac{1}{\hbar}S^{(t,\tau)}_{\rm sc}(t_n,\tau_n)
=
n\omega t_n
+
\frac{i\Gamma+\Delta}{\hbar\omega}\omega\tau_n
-
\frac{U_{\rm p}}{24\hbar\omega}\notag\\
&
(\omega\tau_n)^3[
\frac{(\omega\tau_n)^2}{15}
+(\omega \tilde{t}'_n+\omega\tilde{t}_n)^2
-\frac{(\omega\tau_n)^2}{15}(\omega \tilde{t}'_n+\omega\tilde{t}_n)^2\notag\\
&
-{12}(\omega \tilde{t}'_n+\omega\tilde{t}_n)^4
+\frac{(\omega\tau_n)^5}{420}
],
\end{align}
Using the identities $\zeta_n^2-\zeta_0^2=n$ and $\zeta^2_0={(i\Gamma+\Delta)}/{(\hbar\omega)}$ and the solutions of $t'_n$ and $t_n$ up to the order of $({\hbar\omega}/{U_{\rm p}})^{3/4}$, we arrive at a form of the semiclassical action up to the order of $({\hbar\omega}/{U_{\rm p}})^{3/4}$,
\begin{align}
\frac{1}{\hbar}S^{(t,\tau)}_{\rm sc}(t_n,\tau_n)
=
&
q_{1/4}(n,\frac{i\Gamma+\Delta}{\hbar\omega})(\frac{\hbar\omega}{U_{\rm p}})^{1/4}\notag\\
&
+
q_{3/4}(n,\frac{i\Gamma+\Delta}{\hbar\omega})(\frac{\hbar\omega}{U_{\rm p}})^{3/4},
\end{align}
where
\begin{align}
q_{3/4}(n, \frac{i\Gamma+\Delta}{\hbar\omega}) &
=
(\frac{1}{18})^{1/4}
\frac{1}{1260\sqrt{\zeta_n - \zeta_0}}[103(\zeta_n^2-\zeta_0^2)^2
\notag\\
&
+232\zeta_0\zeta_n(\zeta^2_0+\zeta^2_n)
-184\zeta_0^2\zeta^2_n].
\end{align}

\section{Supplementary figures for the accuracy analysis}\label{APP:supplement_figures}

\begin{figure}
	\includegraphics[width=0.5\textwidth]{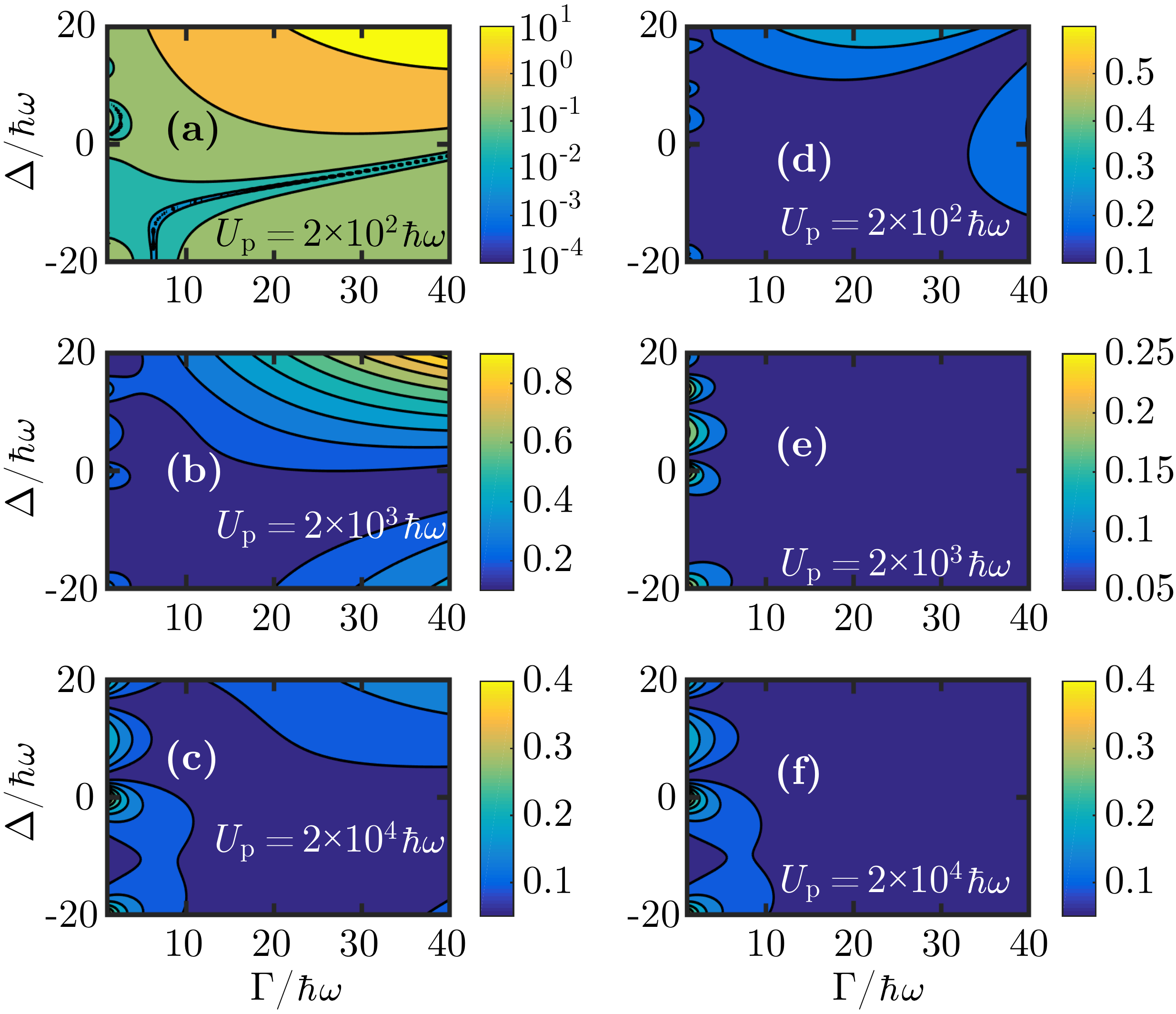}
	\caption{The accuracy of the linear-in-time approximation for the absolute values of the dimensionless sideband amplitudes $Q_{20}$ with varing dephasing and detuning. Left (Right) column: the relative errors in $|Q_{20}|$ without (with) a higher-order correction. The values of the ponderomotive energy $U_{\rm p}$ are chosen as $2\times10^2\hbar\omega$ ((a) and (d)), $2\times10^3\hbar\omega$ ((b) and (e)), and $2\times10^4\hbar\omega$ ((c) and (f)). {\bf The dimension of the momentum space is one (D=1)}.}
	\label{FIG:detuning_abs_n20}
\end{figure}

\begin{figure}
	\includegraphics[width=0.5\textwidth]{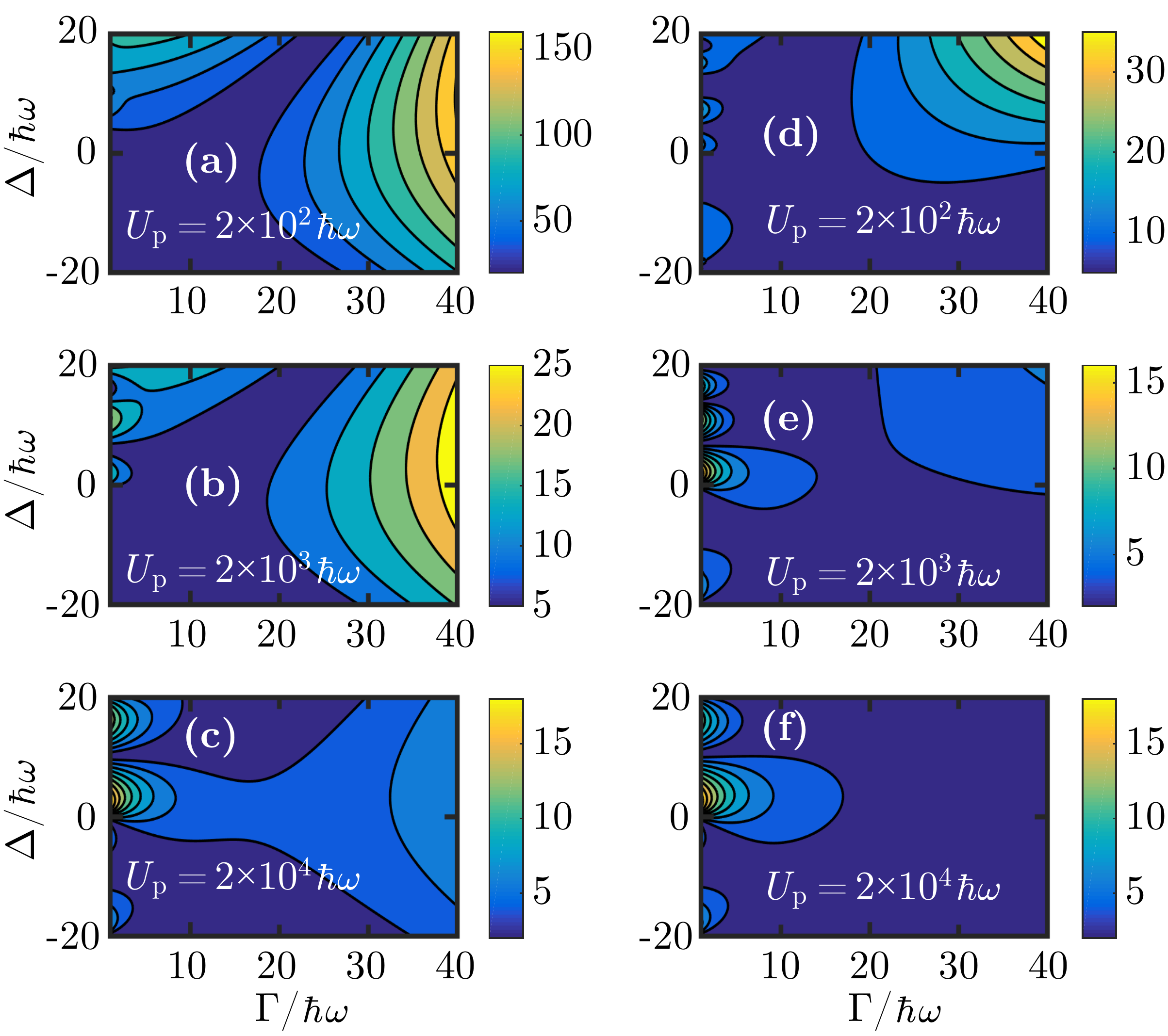}
	\caption{The accuracy of the linear-in-time approximation for the phases of the dimensionless sideband amplitudes $Q_{20}$ with varing dephasing and detuning. Left (Right) column: the absolute errors in the phases of $Q_{20}$ without (with) a higher-order correction. The values of the ponderomotive energy $U_{\rm p}$ are chosen as $2\times10^2\hbar\omega$ ((a) and (d)), $2\times10^3\hbar\omega$ ((b) and (e)), and $2\times10^4\hbar\omega$ ((c) and (f)). {\bf The dimension of the momentum space is one (D=1)}.}
	\label{FIG:detuning_phase_n20}
\end{figure}

\begin{figure}
	\includegraphics[width=0.5\textwidth]{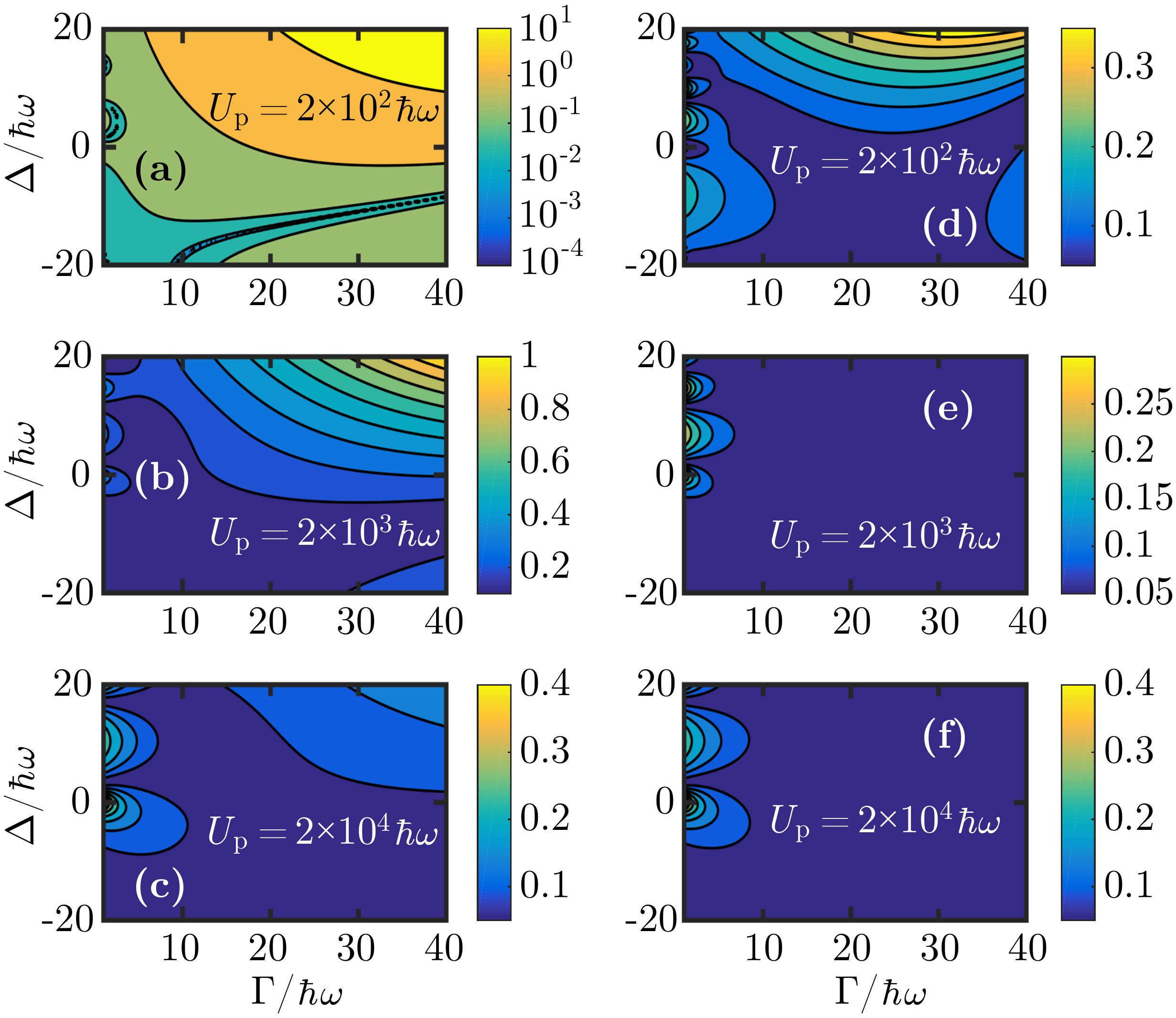}
	\caption{The accuracy of the linear-in-time approximation for the absolute values of the dimensionless sideband amplitudes $Q_{30}$ with varing dephasing and detuning. Left (Right) column: the relative errors in $|Q_{30}|$ without (with) a higher-order correction. The values of the ponderomotive energy $U_{\rm p}$ are chosen as $2\times10^2\hbar\omega$ ((a) and (d)), $2\times10^3\hbar\omega$ ((b) and (e)), and $2\times10^4\hbar\omega$ ((c) and (f)). {\bf The dimension of the momentum space is one (D=1)}.}
	\label{FIG:detuning_abs_n30}
\end{figure}

\begin{figure}
	\includegraphics[width=0.5\textwidth]{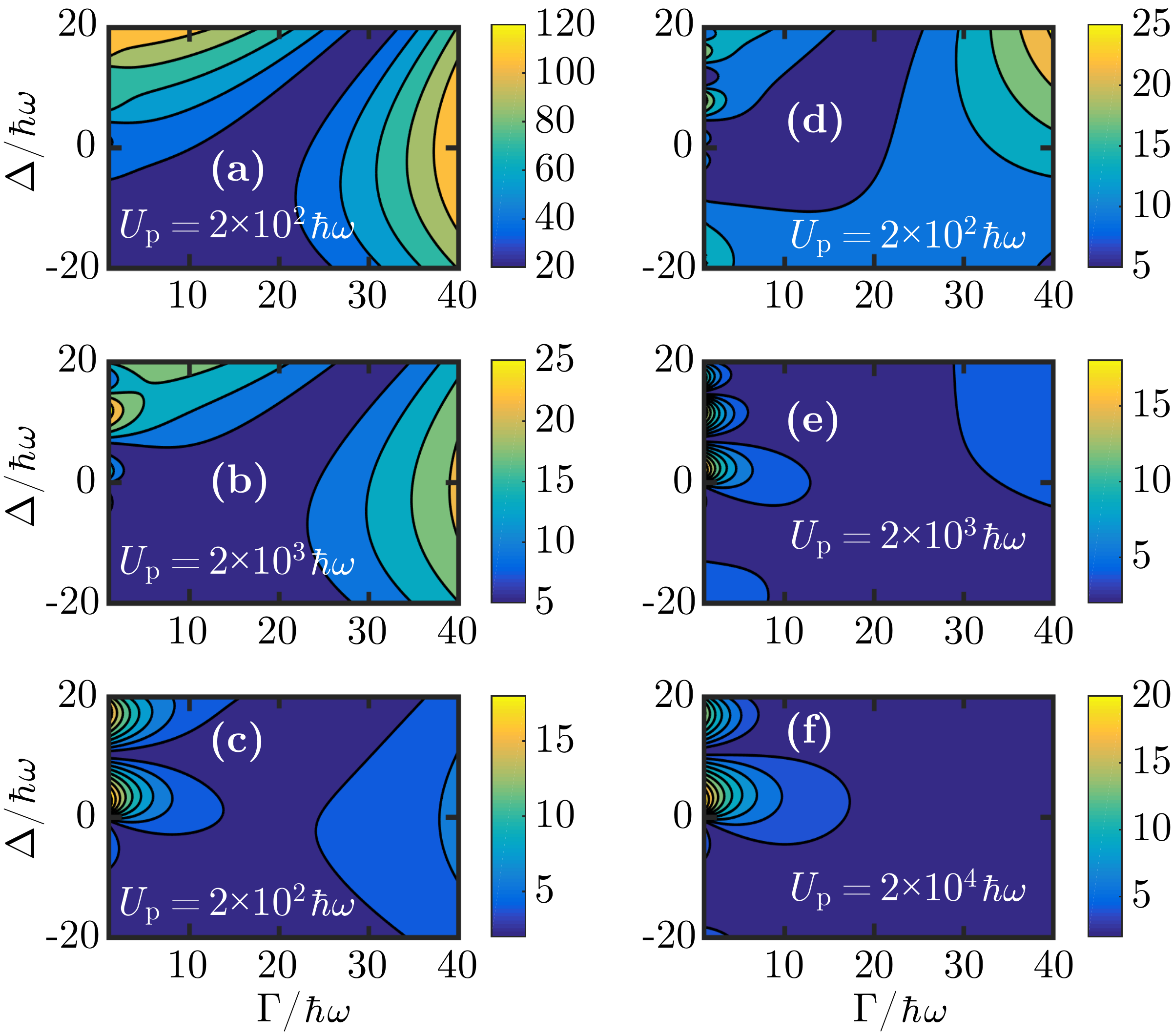}
	\caption{The accuracy of the linear-in-time approximation for the phases of the dimensionless sideband amplitudes $Q_{30}$ with varing dephasing and detuning. Left (Right) column: the absolute errors in the phases of $Q_{30}$ without (with) a higher-order correction. The values of the ponderomotive energy $U_{\rm p}$ are chosen as $2\times10^2\hbar\omega$ ((a) and (d)), $2\times10^3\hbar\omega$ ((b) and (e)), and $2\times10^4\hbar\omega$ ((c) and (f)). {\bf The dimension of the momentum space is one (D=1)}.}
	\label{FIG:detuning_phase_n30}
\end{figure}

\begin{figure}
	\includegraphics[width=0.5\textwidth]{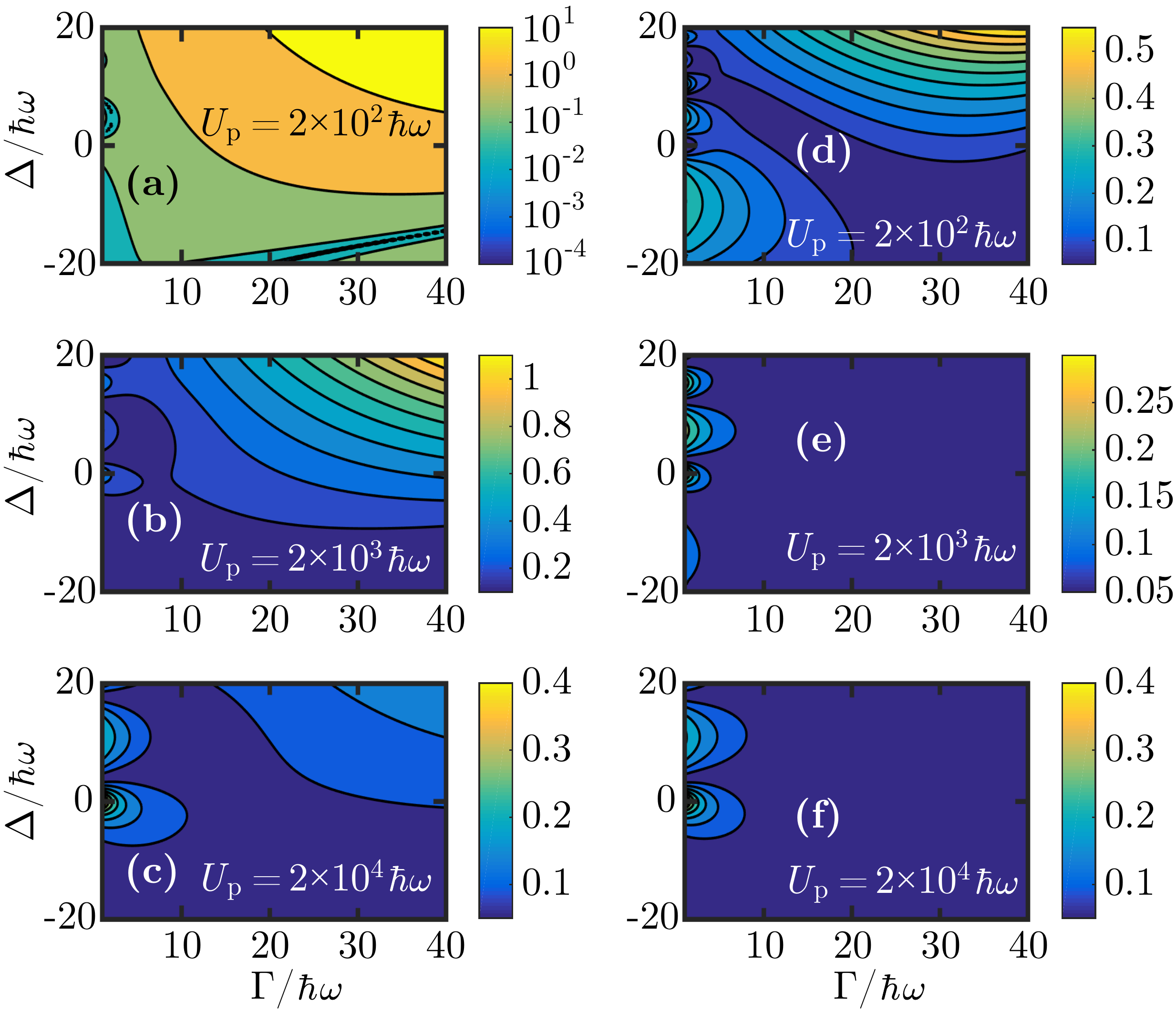}
	\caption{The accuracy of the linear-in-time approximation for the absolute values of the dimensionless sideband amplitudes $Q_{40}$ with varing dephasing and detuning. Left (Right) column: the relative errors in $|Q_{40}|$ without (with) a higher-order correction. The values of the ponderomotive energy $U_{\rm p}$ are chosen as $2\times10^2\hbar\omega$ ((a) and (d)), $2\times10^3\hbar\omega$ ((b) and (e)), and $2\times10^4\hbar\omega$ ((c) and (f)). {\bf The dimension of the momentum space is two (D=2)}.}
	\label{FIG:detuning_abs_n40_dimk2}
\end{figure}

\begin{figure}
	\includegraphics[width=0.5\textwidth]{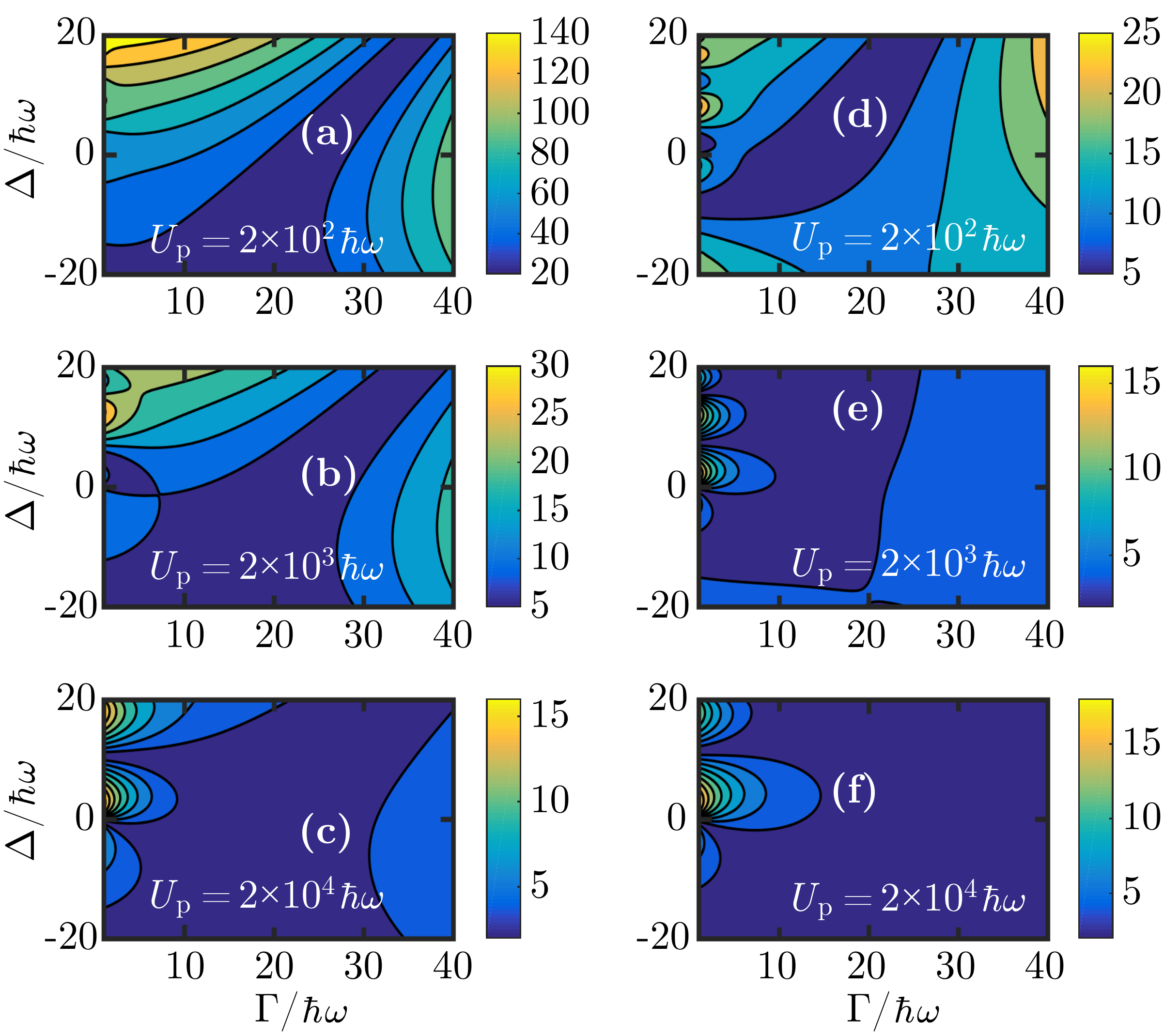}
	\caption{The accuracy of the linear-in-time approximation for the absolute values of the dimensionless sideband amplitudes $Q_{40}$ with varing dephasing and detuning. Left (Right) column: the relative errors in $|Q_{40}|$ without (with) a higher-order correction. The values of the ponderomotive energy $U_{\rm p}$ are chosen as $2\times10^2\hbar\omega$ ((a) and (d)), $2\times10^3\hbar\omega$ ((b) and (e)), and $2\times10^4\hbar\omega$ ((c) and (f)). {\bf The dimension of the momentum space is two (D=2)}.}
	\label{FIG:detuning_phase_n40_dimk2}
\end{figure}

\begin{figure}
	\includegraphics[width=0.5\textwidth]{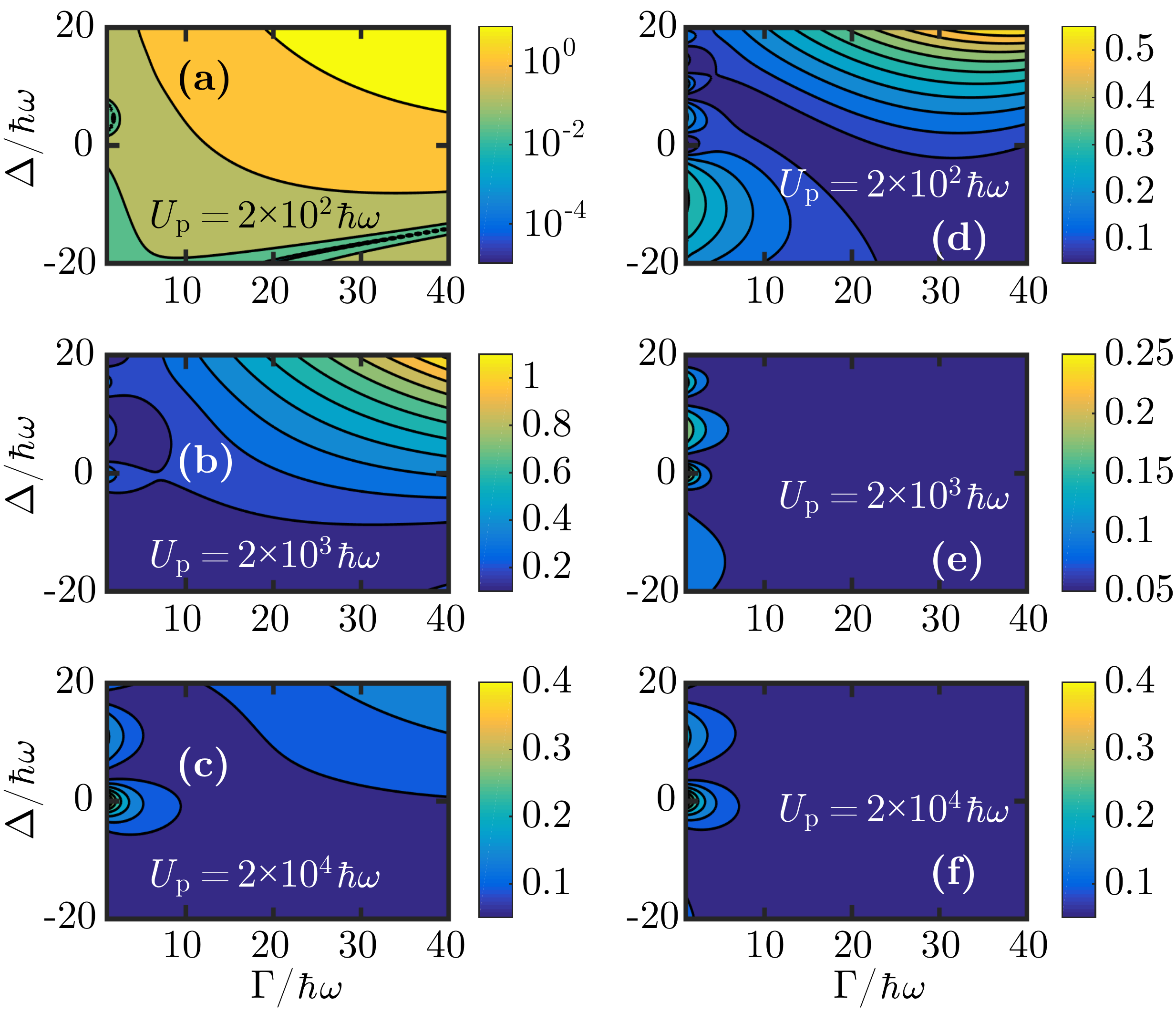}
	\caption{The accuracy of the linear-in-time approximation for the absolute values of the dimensionless sideband amplitudes $Q_{40}$ with varing dephasing and detuning. Left (Right) column: the relative errors in $|Q_{40}|$ without (with) a higher-order correction. The values of the ponderomotive energy $U_{\rm p}$ are chosen as $2\times10^2\hbar\omega$ ((a) and (d)), $2\times10^3\hbar\omega$ ((b) and (e)), and $2\times10^4\hbar\omega$ ((c) and (f)). {\bf The dimension of the momentum space is three (D=3)}.}
	\label{FIG:detuning_abs_n40_dimk3}
\end{figure}

\begin{figure}
	\includegraphics[width=0.5\textwidth]{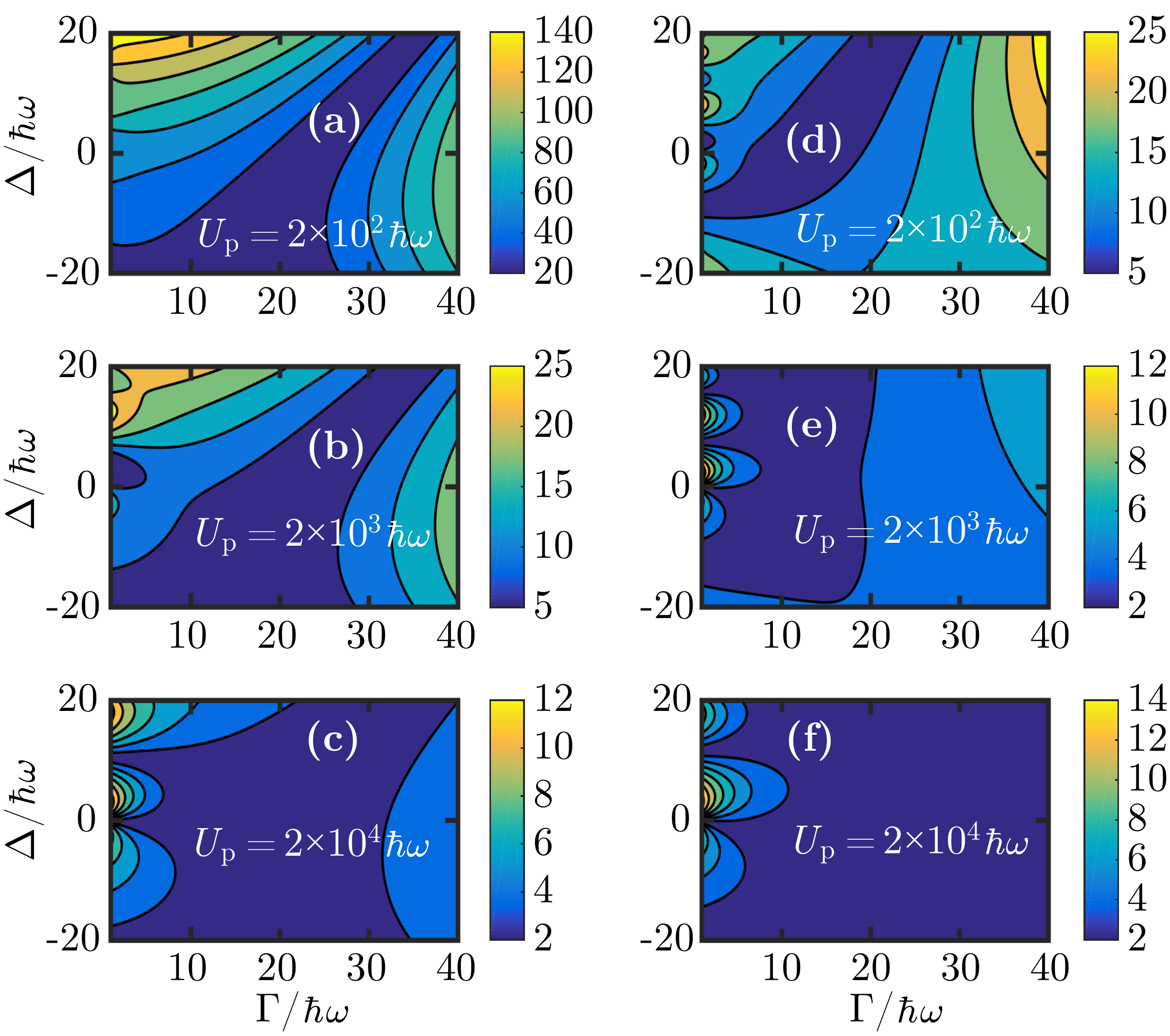}
	\caption{The accuracy of the linear-in-time approximation for the absolute values of the dimensionless sideband amplitudes $Q_{40}$ with varing dephasing and detuning. Left (Right) column: the relative errors in $|Q_{40}|$ without (with) a higher-order correction. The values of the ponderomotive energy $U_{\rm p}$ are chosen as $2\times10^2\hbar\omega$ ((a) and (d)), $2\times10^3\hbar\omega$ ((b) and (e)), and $2\times10^4\hbar\omega$ ((c) and (f)). {\bf The dimension of the momentum space is three (D=3)}.}
	\label{FIG:detuning_phase_n40_dimk3}
\end{figure}

%

\end{document}